%% file: main.tex
\RequirePackage{fix-cm}
\documentclass[smallextended]{svjour3}       % onecolumn (second format)
\smartqed  % flush right qed marks, e.g. at end of proof
\usepackage{graphicx}
\usepackage{url}
\usepackage{subfigure}
\usepackage{tcolorbox}
\usepackage{multirow}
\usepackage{amssymb,amsfonts}
\usepackage{pifont}%for \xmark and \cmark
\usepackage{xcolor,colortbl} %for \rowcolor{}
\usepackage{listings}
\usepackage{hyperref}
\input{commands}

\begin{document}

\title{Development of \rs for software engineering: the 
\projectName experience%\thanks{Grants or other 
%notes
}

\titlerunning{Development of RSSE: the \CM experience}        % if too long for running head

\author{Juri Di Rocco \and Davide Di Ruscio \and Claudio Di Sipio
\and Phuong T. Nguyen \and Riccardo Rubei}


\institute{Juri Di Rocco, Davide Di Ruscio, Phuong T. Nguyen, Claudio Di Sipio, 
Riccardo Rubei \at
		DISIM - University of L'Aquila (Italy)\\
	\email{firstname.lastname@univaq.it}           %  \\
}

\maketitle

\begin{abstract}
		%%%% ALL OF THIS NEEDS TO BE REWRITTEN/REWORDED. IT COMES FROM 
	%%%%OTHER PAPERS
	To perform their daily tasks, developers intensively make use of existing resources by consulting open source software 
	(OSS) repositories. Such platforms contain rich data  sources, \eg code snippets, documentations, and user discussions, that can be 
	useful for supporting development activities. Over the last decades, several techniques and tools have been promoted to provide 
	developers with innovative features, aiming to bring in improvements in terms of development effort, cost savings, and 
	productivity.
	In the context of the EU H2020 CROSSMINER project, a set of recommendation 
	systems has been conceived to assist software 
	programmers in different phases of the development process. The 
	systems provide developers with various artifacts, such as third-party 
	libraries, documentation about how to use the APIs being adopted, or 
	relevant API function calls. To develop such recommendations, various 
	technical choices have been made to overcome issues related to several 
	aspects including the lack of baselines, limited data availability, 
	decisions about the performance measures, and evaluation approaches. 
	
	This paper is an experience report to present the knowledge pertinent to the set of recommendation systems developed 
	through the CROSSMINER project. We explain in detail the challenges we had to deal with, together with the related lessons learned 
	when developing and evaluating these systems. Our aim is to provide the research community with concrete takeaway messages that are 
	expected to be useful for those who want to develop or customize their own recommendation systems. The reported experiences can 
	facilitate interesting discussions and research work, which in the end contribute to the advancement of recommendation systems 
	applied to solve different issues in Software Engineering. 
	
%	start to build 
%	 one can refer to when
%	First, we listed the challenges that we faced when designing and implementing the CROSSMINER recommendation systems. Second, the lessons are presented as the experience gained by overcoming the challenges. In this way, we hope that the lessons are the takeaway messages that the research community can use to build their own recommendation systems. 
	
	% that have been conceived in the context of the
	
%	the experiences gained by executing a research project. We describe 
	
	%	Moreover, we describe also our planned medium and long-term objectives 
	%in 
	%	the same application domain.
	

\end{abstract}

	\section{Introduction}
	\label{sec:Introduction}
	\input{src/Introduction}

%	\setcounter{tocdepth}{2}
%	\tableofcontents
	
%	\section{Background}
%	\label{sec:Motivation}
%	\input{src/Context}
	
	\section{The CROSSMINER project}
	\label{sec:CROSSMINER}
	\input{src/Crossminer}

	\section{Challenges and lessons learned from eliciting the requirements of the \projectName \rs} 
	\label{sec:requirements}

\input{src/requirements}

	\section{Challenges and lessons learned from developing the \projectName \rs}	
	\label{sec:DevelopingRs}
	\input{src/Developing}

	\section{Challenges and lessons learned from the evaluation of the \projectName \rs}
	\label{sec:evaluation}
	\input{src/evaluation}

%	\section{Lessons learned}
%	\label{sec:Lessons}
%	\input{src/Lessons}

%	\section{Assessing the quality of recommendation system} 
%	\label{sec:Evaluation}
		
	\section{Related work}
	\label{sec:RelatedWork}
	\input{src/RelatedWork.tex}

	\section{Conclusion and future work}
	\label{sec:Conclusions}
	\input{src/Conclusions}

	\begin{acknowledgements}
%		The research described in this paper has been carried out as part of the
%		CROSSMINER Project, which has received funding from the European Union's
%		Horizon 2020 Research and Innovation Programme under Grant 732223
		The research described in this paper has been carried out as part of the CROSSMINER Project, which has received funding from the European Union's Horizon 2020 Research and Innovation Programme under Grant 732223. We thank the anonymous reviewers for their valuable comments and suggestions that helped us improve the paper.
	\end{acknowledgements}


	\bibliographystyle{spmpsci}      % mathematics and physical sciences

\bibliography{main}
\end{document}

%% file: commands.tex
\usepackage{xspace}
\usepackage{amssymb}
\usepackage[colorinlistoftodos]{todonotes}
\usepackage{xcolor}

\newcommand*{\ie}{i.e.,\@\xspace}
\newcommand*{\eg}{e.g.,\@\xspace}

\newcommand*{\RM}{README\@\xspace}

\newcommand*{\GH}{GitHub\@\xspace}
\newcommand*{\PR}{PROMPTER\@\xspace}
\newcommand*{\CS}{CrossSim\@\xspace}
\newcommand*{\CR}{CrossRec\@\xspace}
\newcommand*{\FC}{FOCUS\@\xspace}

\newcommand*{\CM}{CROSSMINER\@\xspace}

\makeatletter
\newcommand*{\etc}{%
	\@ifnextchar{.}%
	{etc}%
	{etc.\@\xspace}%
}
\makeatother
\newcommand*{\etal}{\emph{et~al.}\@\xspace}
\newcommand{\cmark}{\ding{51}}%

\newcommand*\circled[1]{\tikz[baseline=(char.base)]{\color{black} 
		\node[shape=circle,draw=cyan,fill=black!10!white,inner sep=.3pt] (char) 
		{{{\texttt\textbf #1}}};}}

\newcommand{\code}[1]{{\texttt{#1}}}

\newcommand{\projectName}{CROSSMINER\@\xspace}

\newcommand{\bnetwork}{MNBN}

\newcommand*{\rs}{recommendation systems\@\xspace}

\newcommand{\challenge}[2]{
	\begin{tcolorbox}[boxrule=0.5pt]
		{\bfseries #1}: {#2}
	\end{tcolorbox}
}

\definecolor{darkgray}{gray}{0.78}
\definecolor{lightgray}{gray}{0.85}
\definecolor{verylightgray}{gray}{0.95}
\definecolor{codegreen}{rgb}{0,0.6,0}
\definecolor{codegray}{rgb}{0.5,0.5,0.5}
\definecolor{codepurple}{rgb}{0.58,0,0.82}
\definecolor{backcolour}{rgb}{0.95,0.95,0.92}
\lstdefinestyle{java}{
	backgroundcolor=\color{backcolour},   commentstyle=\color{codegreen},
	keywordstyle=\color{magenta},
	numberstyle=\tiny\color{codegray},
	stringstyle=\color{codepurple},
	basicstyle=\ttfamily\scriptsize,
	breakatwhitespace=false,         
	breaklines=true,                 
	captionpos=b,                    
	keepspaces=true,                 
	numbers=left,                    
	numbersep=5pt,                  
	showspaces=false,                
	showstringspaces=false,
	showtabs=false,                  
	tabsize=2
}

%% file: src/Introduction.tex
%In the context of o
%high quality 

Open-source software (OSS) forges, such as \GH or Maven, offer many software projects that deliver stable and well-documented 
products. Most OSS forges typically sustain vibrant user and expert communities which in turn provide decent support, both for 
answering user questions and repairing reported software bugs. Moreover, OSS platforms are also an essential source of consultation for 
developers in their daily development tasks~\cite{7887704}. Code reusing is an intrinsic feature of OSS, and developing new software by 
leveraging existing open source components allows one to considerably reduce their development effort. %, and thus being beneficial to 
%the overall software life cycle.
The benefits resulting from the reuse of properly selected open-source projects are manifold including the fact that the system being 
implemented relies on open source code, \textit{``which is of higher quality than the custom-developed code's first 
incarnation''}~\cite{SS04}. In addition to source code, also metadata available from different related sources, \eg communication 
channels and bug tracking systems, can be beneficial to the development life cycle if being properly mined 
\cite{DBLP:journals/ese/PonzanelliBPOL16}. Nevertheless, given a plethora of data sources, developers would struggle to look 
for and approach the sources that meet their need without being equipped with suitable machinery. %, in the hope of transforming them 
%to practical knowledge. 
Such a process is time-consuming since the problem is not a lack, but in contrast, an overload of information coming from heterogeneous 
and rapidly evolving sources. In particular, when developers join a new project, they have to typically master a considerable number of 
information sources (often at a short time)~\cite{Dagenais:2010:MNS:1806799.1806842}. In this respect, the deployment of systems that 
use existing data to improve developers' experience is of paramount importance.

%The introduction of \rs to the domain of software development brings 
%substantial benefits. Among others, r
%, there is still the need to make them more effective and efficient
%In recent years, considerable effort has been made to provide automated assistance to developers in navigating large information spaces and giving recommendations. 
% that meet their taste and expectation

Recommendation systems are a crucial component of several online shopping systems, allowing business owners to offer 
personalized products to customers \cite{Linden:2003:ARI:642462.642471}. The development of such systems has culminated in well-defined 
recommendation algorithms, which in turn prove their usefulness in other fields, such as entertainment industry 
\cite{Gomez-Uribe:2015:NRS:2869770.2843948}, or employment-oriented service \cite{DBLP:conf/recsys/WuSCTP14}. Recommendation systems in 
software engineering~\cite{robillard_recommendation_2014} (RSSE hereafter) have been conceptualized on a comparable basis, \ie they 
assist developers in navigating large information spaces and getting instant recommendations that are helpful to solve a particular 
development task~\cite{Nguyen:2019:FRS:3339505.3339636,DBLP:journals/ese/PonzanelliBPOL16}. In this sense, RSSE provide developers with 
useful recommendations, which may consist of different items, such as code 
examples~\cite{Fowkes:2016:PPA:2950290.2950319,Moreno:2015:IUT:2818754.2818860,Nguyen:2019:FRS:3339505.3339636}, 
topics~\cite{10.1145/3382494.3410690,10.1145/3383219.3383227}, %reusable source code, 
third-party components~\cite{Nguyen:2019:JSS:CrossRec,6671293}, documentation~\cite{ponzanelli_prompter:_2016,RUBEI2020106367}, to name 
a few. %, and gained momentum in recent years

%, however though a lot of improvements have been obtained so far

While the issue of designing and implementing generic \rs has been carefully addressed by state-of-the-art 
	studies~\cite{tubiblio77729,robillard_recommendation_2014}, there is a lack of proper references for the design of a 
	recommendation system in a concrete context, \ie satisfying requirements by various industrial partners. By means of a thorough 
	investigation of the related work, we realized that existing studies tackled the issue of designing and implementing a 
	recommendation system for software engineering in general. However, to the best of our knowledge, an experience report extracted 
	from real development projects is still missing. The report presented in this paper would come in handy for those who want to 
	conceive or customize their recommendation systems in a specific context. For example, a developer may be interested in 
	understanding which techniques are suitable for producing recommendations; or how to capture the developer’s context; or which is 
	the most feasible way to present recommendation outcomes, to name a few.
	
In the context of the EU CROSSMINER project\footnote{\url{https://www.crossminer.org}} we exploited cutting-edge information 
	retrieval techniques to build \rs, providing software developers with practical advice	on various tasks through an Eclipse-based 
	IDE and dedicated analytical Web-based dashboards. Based on the project's mining tools, developers can select open-source software 
	and get real-time recommendations while working on their development tasks. This paper presents an experience report pertaining to the implementation of the \CM \rs, with focus on three main phases, \ie \emph{Requirements elicitation}, \emph{Development}, and \emph{Evaluation}. We enumerate the challenges that we faced when working with 
these phases and present the lessons gained by overcoming the challenges. With this work, we aim at providing the research community at large with practical takeaway messages that one can consult when building their recommendation systems.

% Interestingly, little attention has been paid to transfer and adapt such techniques in mining OSS repositories. %Though remarkable progress can be seen in such a direction, there is still room for improvement. 
%There is a lack of a proper scheme that facilitates a unified consideration of various OSS artifacts and recommendations: most of the existing approaches consider the constituent components of the OSS ecosystem separately, without paying much attention to their mutual connections.

%the research community
%the lessons are the 
%This is the main motivation for the current paper: We fill the gap by presenting. 
%In this respect, it is crucial to. 
%there is still the need to present 
%Research Gap (\textbf{Comment 1.2}):
%on the realization of the \CM \rs by concentrating on three main phas\-es 

%This aims at creating a benchmark that can be widely used by the community.

\vspace{.2cm}

%are presented as the experience 
% More importantly, we identify the issues and the ways we tackled them by distinguishing. 
%First, we enumerate a list of challenges and lessons learned from the project. 
%In particular, we discuss the challenges as well as 
%the lessons learned from the CROSSMINER experience. 

\noindent
\textbf{Outline of the paper:} The paper is structured as follows: 
%Section \ref{sec:Motivation} shows the motivation and the background %of this %%work. 
	Section \ref{sec:CROSSMINER} gives an overview of the CROSSMINER project and the underpinning motivations. 
	Sections~\ref{sec:requirements}--\ref{sec:evaluation} discuss the challenges we had to address while conceiving the \projectName \rs and the corresponding lessons learned that we would like to share with the community as well as with potential developers of new \rs.
	Section \ref{sec:RelatedWork} reviews the related work and finally, Section \ref{sec:Conclusions} sketches perspective work and concludes the paper.
%RelatedWork

%% file: src/Crossminer.tex
In recent years, software development activity has reached a high
degree of complexity, led by the heterogeneity of the components,
data sources and tasks.
%To address such challenges, suitable machinery is needed to transform raw data into practical knowledge that can really help 
%developers with their programming tasks. 
The 
adoption of \rs in software engineering (RSSE)
aims at supporting developers in navigating large information
spaces and getting instant suggestions that might be helpful
to solve a particular development task~\cite{robillard_recommendation_2014}.

%Several RSSEs are nowadays available to support software development 
%activities by using well-defined strategies. In the following, we 
%make an overview of existing approaches by focusing on those that 
%are 
%mainly related to the CROSSMINER \rs, which are in focus of this 
%paper. In particular, source-code search techniques, source-code 
%recommenders, and API documentation and Q\&A recommenders  

In the context of open-source software, developing new software systems by reusing existing open-source components raises relevant 
challenges related to at least the following activities 
\cite{Karlsson1995}: \textit{(i)} searching for candidate components, 
\textit{(ii)} evaluating a set of retrieved candidate components to 
find the most suitable ones, and \textit{(iii)} adapting the selected 
components to fit some specific requirements. The \CM project conceived techniques and tools for extracting knowledge from existing 
open source components and use it to provide developers with real-time recommendations that are relevant to the current 
development task. 
\begin{figure}[t!]
	\centering
	\includegraphics[width=0.9\linewidth]{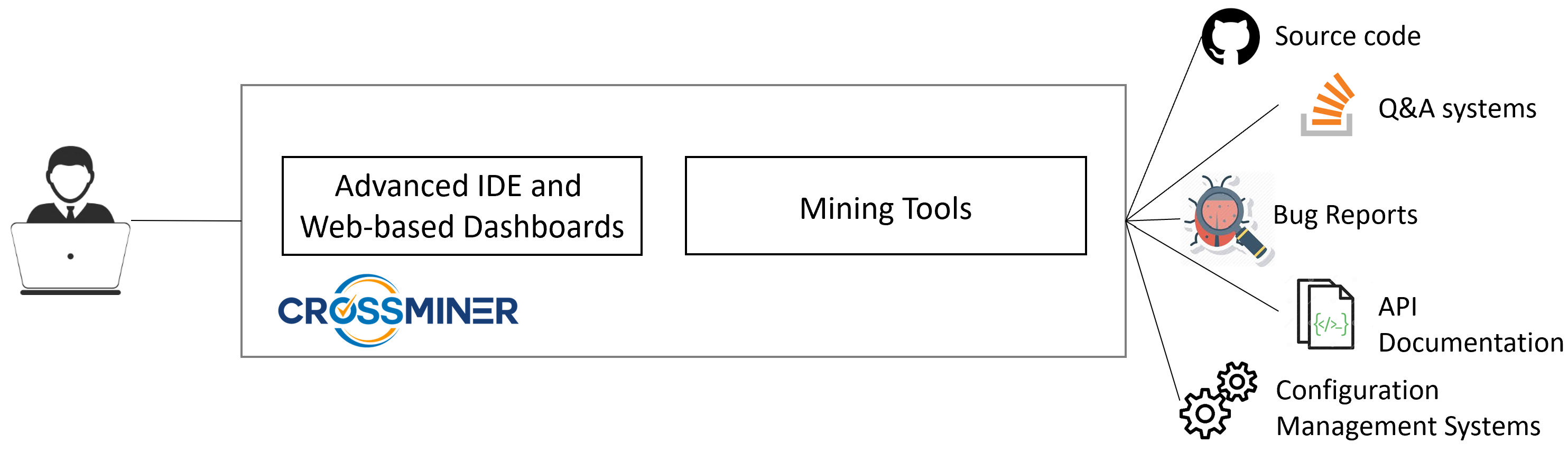}
	\caption{CROSSMINER Overview.}
	\label{fig:crossminer-overview}
\end{figure}

As shown in Fig. \ref{fig:crossminer-overview}, the \projectName components are
conceptually in between the developer and all the different and heterogeneous 
data sources (including source code, bug tracking systems, and communication 
channels) that one needs to interact with when understanding and using existing 
open-source components. In particular, an Eclipse-based IDE and Web-based 
dashboards make use of data produced by the mining tools working on the 
the back-end of the \projectName infrastructure to help developers perform the 
current development tasks. \CM is under the umbrella of Eclipse Research Labs 
with the name Eclipse SCAVA.\footnote{\url{https://www.eclipse.org/scava/}}

%Recommender systems typically implement four main activities i.e., data 
%pre-processing, capturing context, producing recommendations, presenting 
%recommendations. In this respect, both the 

\subsection{\projectName as a set of \rs}

\begin{figure}[t!]
	\centering
	\fbox{\includegraphics[width=.98\textwidth]{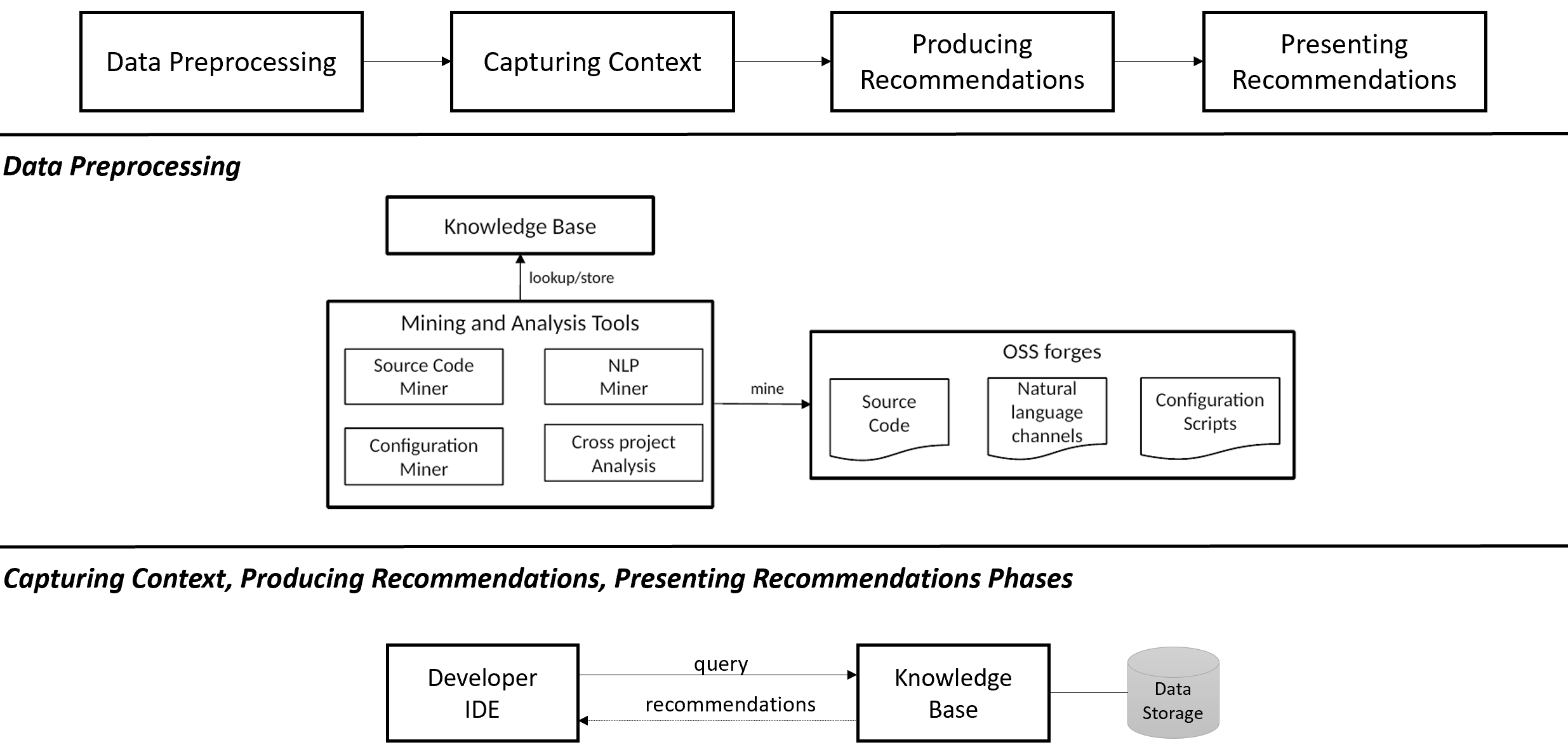}}
	%	\vspace{-.4cm}
	\caption{High-level view of the CROSSMINER project.}
	%	\vspace{-.4cm}
	\label{fig:CROSSMINER}
\end{figure}

Figure \ref{fig:CROSSMINER} shows \projectName from a different 
perspective. In particular, \projectName can be seen as a set of recommendation systems, each designed to implement the four main 
activities, which are typically defined for any \rs, \ie data pre-processing, 
capturing context, producing recommendations, and presenting recommendations \cite{robillard_recommendation_2014} as shown in the upper 
side of Fig.~\ref{fig:CROSSMINER}. Accordingly, the 
\projectName solution is made up of four main modules: the \code{Data Preproces\-sing} module contains tools that extract metadata from 
OSS repositories (see the middle part of Fig.~\ref{fig:CROSSMINER}). Data can be of different types, such as source code, 
configuration, or cross-project relationships. Natural language processing 
(NLP) tools are also deployed to analyze developer forums and discussions. The 
collected data is used to populate a knowledge base which serves as the core 
for the mining functionalities. By capturing developers' activities 
(\code{Capturing Context}), an IDE is able to generate and display 
recommendations (\code{Producing Recommendations} and \code{Presenting 
	Recommendations}). In particular, the developer context is used as a query sent 
to the knowledge base that answers with recommendations that are relevant to the developer contexts (see the lower side of 
Fig.~\ref{fig:CROSSMINER}). Machine learning techniques are used to 
infer knowledge underpinning the creation of relevant real-time 
recommendations.

The \projectName~ \code{knowledge base} allows developers to gain insights into 
raw data 
produced by 
different mining tools, which are the following ones:

\begin{itemize}
	
	\item \textit{Source code miners} to extract and store actionable knowledge 
	from the source code of a collection of open-source projects;
	
	\item \textit{NLP miners} to extract quality metrics related to the 
	communication channels, and bug tracking systems of OSS projects by using 
	Natural Language Processing and text mining techniques;
	
	\item \textit{Configuration miners} to gather and analyze system 
	configuration artefacts and data to provide an integrated DevOps-level view 
	of a considered open source project;
	
	\item \textit{Cross-project miners} to infer cross-project relationships 
	and additional knowledge underpinning the provision of real-time 
	recommendations;
	
\end{itemize}

%In the next sections, we make an overview of representative \CM \rs, 
%which are then used to discuss the challenges we faced when developing them 
%and 
%the related lessons learned.

\begin{table}[t!]
	\centering
%	\color{blue}
	\footnotesize
	\scriptsize
	%	\vspace{-.2cm}
	\caption{The \CM use cases.}
	%	\vspace{-.2cm}
	\label{tab:UseCases}
	\begin{tabular}{|p{0.3cm} | p{2.9cm}| p{5cm} | p{1.75cm}|}  
		\hline				
		\textbf{No.} & \textbf{Artifact} & \textbf{Description} & 
		\textbf{Developed tool}\\ \hline
		\circled{1} & \emph{Similar OSS projects} & We crawl data from OSS 
		platforms to find similar projects to the system being developed, with 
		respect to different criteria, \eg external dependencies, or API 
		usage~\cite{Nguyen:2019:JSS:CrossSim}. This type of recommendation is 
		beneficial to the development since it helps developer learn how 
		similar projects are implemented. & \CS \\ \hline		
		\circled{2} & \emph{Additional components} & During the development 
		phase, programmers search for components that projects similar to the 
		one under developed have also included, for instance, a list of 
		external libraries \cite{McMillan:2010:RSC:1808920.1808925}, or API 
		function calls~\cite{Moreno:2015:IUT:2818754.2818860}. & \CR, \FC \\ 
		\hline			
		\circled{3} & \emph{Code snippets}  & Well-defined snippets showing how 
		an API is used in practice are extremely useful. These snippets provide 
		developers with a deeper insight into the usage of the APIs being 
		included~\cite{Nguyen:2019:FRS:3339505.3339636}. & \FC \\ \hline	
%		\circled{4} & \emph{Documentations} & These are technical documents, 
%		tutorials, communication channels, etc., that are relevant to the code 
%		being developed, for instance by mining external experiences from Stack 
%		Oveflow. & \textbf{MAYBE TO BE REMOVED}\\ \hline	
%		\circled{5} & \emph{API changes and consequences} & Changes of 
%		libraries will have a certain effect on the depending projects. It is 
%		necessary to notify developers and recommend amendments to preserve 
%		program compatibility. & \textbf{MAYBE TO BE REMOVED}\\ \hline	
		\circled{4} & \emph{Relevant topics} & \GH uses tags as a means to 
		narrow down the search scope. The goal is to help developers approach 
		repositories, and thus increasing the possibility of contributing to 
		their development and widespread their usage. & \bnetwork \\ \hline	
	\end{tabular}
	\vspace{-.2cm}
%	\color{black}
\end{table}

The \CM \rs have been developed to satisfy the requirements by six 
industrial use-case partners of the project working on different domains 	
including  IoT, multi-sector IT services, API co-evolution, software analytics, 
software quality assurance, and OSS 
forges.\footnote{\url{https://www.crossminer.org/consortium}} In particular, 
Table~\ref{tab:UseCases} specifies the main use cases solicited by our 
industrial partners. To satisfy the given requirements the following 
\rs have been developed:

	\begin{itemize}
%		\color{blue}		
		\item \textit{\CS}~\cite{8498236,Nguyen:2019:JSS:CrossSim} -- It is an 
		approach for 
		recommending similar projects with respect to the 
		third-party library usage, stargazers and commiters, given a specific 
		project;
		
		\item \textit{\CR}~\cite{Nguyen:2019:JSS:CrossRec} -- It is a framework 
		that makes use of
		{\bf C}ross Projects {\bf R}elation\-ships among {\bf O}pen {\bf 
		S}ource 
		{\bf S}oftware Repositories to build a library {\bf Rec}om\-mendation 
		System on top of \CS;
		
		\item \textit{\FC}~\cite{Nguyen:2019:FRS:3339505.3339636,9359479} -- The system assists developers by providing them with API function calls and source code snippets that are 
		relevant for the current development context;
		
		\item \textit{\bnetwork}~\cite{10.1145/3383219.3383227} --  It is an 
		approach based on a Multinomial Naive Bayesian network technique to 
		automatically recommend topics given the \RM file(s) of an input 
		repository.
	\end{itemize}

By referring to Fig. \ref{fig:CROSSMINER}, the developed \rs are 
implemented in the \code{Knowledge Base} component. Moreover, it is important to remark that even though such tools can be used in an 
integrated manner directly from the \code{Developer IDE}, their combined usage is not mandatory. They are different services that 
developers can even use separately according to their needs.

For more details about the \rs developed in the context of the \projectName projects, readers can refer to the related papers 
presenting them. Without giving details on each tool's inner technicalities, in the following sections, we focus on the challenges we 
faced while conceiving the \projectName \rs, and on the lessons that we learned on the way. We believe that sharing them with the 
community is desirable because of two main motivations:
	
\begin{itemize}
	\item The \rs have been developed in a real context to cope with 
	industrial 	needs of different use-case partners;

	\item According to the evaluation procedure performed towards the 
	end of 	the \projectName project, the industrial partners have been 
	particularly 
	satisfied by the developed \rs, which have been mainly graded as 
	\textit{excellent} by most of the partners that were asked to express their 
	judgement in the range \textit{insufficient}, \textit{sufficient}, 
	\textit{good}, \textit{excellent} (see the public deliverable D8.16\footnote{\projectName D8.16 Case Study Evaluation - 
	\url{https://cordis.europa.eu/project/id/732223/results}} for more 
	details).
	
\end{itemize}

	%\subsubsection{\CS~-- Recommending project alternatives with similar APIs}
	%\input{src/CrossSim2}
	%%\paragraph{Motivating scenario}
	%%\paragraph{Approach overview}
	%%\paragraph{Evaluation methodology}
	%
	%\subsubsection{\CR~-- Third-party libraries recommendation}
	%\input{src/CrossRec}
	%%\paragraph{Motivating scenario}
	%%\paragraph{Approach overview}
	%%\paragraph{Evaluation methodology}
	%
	%\subsubsection{\FC~-- Suggesting API function calls and code snippets}
	%\input{src/FOCUS}
	%%\paragraph{Motivating scenario}
	%%\paragraph{Approach overview}
	%%\paragraph{Evaluation methodology}
	%
	%
	%
	%\subsubsection{\bnetwork~-- Suggesting GitHub repository topics}
	%\input{src/mbn}

%\todo{Add a paragraph mentioning that the development of the tools have been 
%difficult from different points of view starting from the requirements up  to 
%the evaluation. It is important to do not say/anticipate contents already 
%described at the beginning of the next section. Mention the the use case 
%partners have been satisfied and they are happy about the provided tools. 
%Maybe 
%mention some numbers of the evaluation?}
%

\subsection{The \projectName development process}

\begin{figure}[h!]
	\centering
	\includegraphics[width=0.68\linewidth]{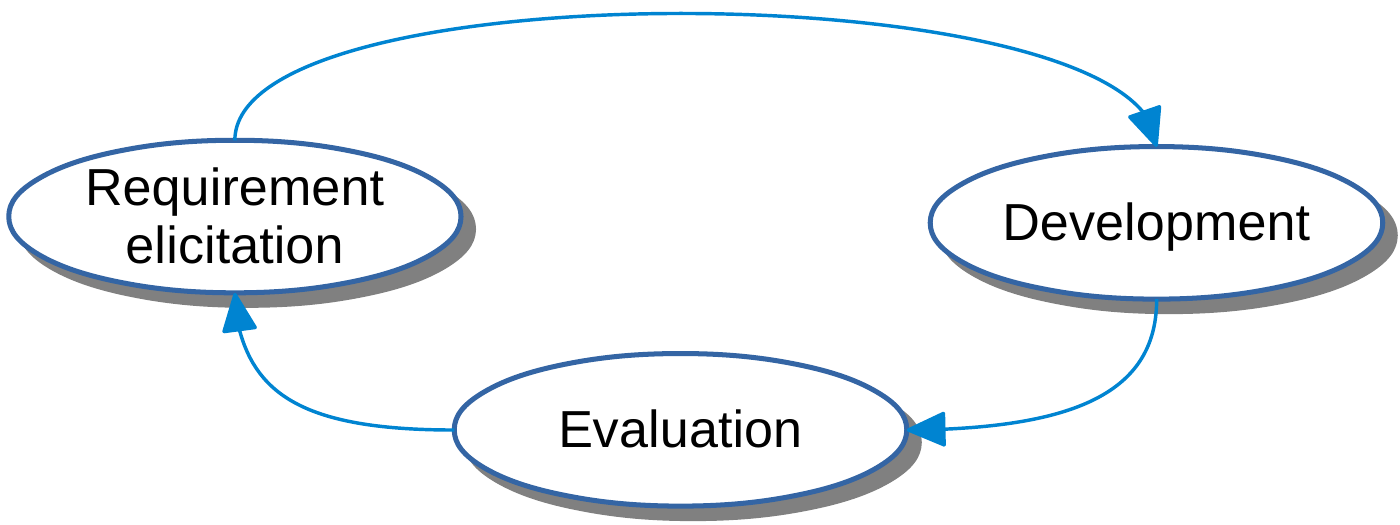}%process-mini
	\caption{Main activities underpinning the development of the \CM \rs.}
	\label{fig:process-mini}
\end{figure}

The development of the \projectName \rs has been done by following an iterative 
process, as shown in Fig.~\ref{fig:process-mini}. In particular,  to produce the 
\rs that are now part of the \projectName platform, the following steps have 
been undertaken: %we had to undertake:

\begin{itemize}
	
	\item \textit{Requirement elicitation}: identification of the expected 
	features provided 
	by the     \projectName platform in terms of recommendations and 
	development support;
	
	\item \textit{Development}: implementation of the needed \rs to accommodate 
	the requirements defined in the previous step;
	
	\item \textit{Evaluation:} assess the performance of the produced 
	recommendations by using properly defined evaluation procedures and 
	selected metrics.
	
\end{itemize}

\vspace{-.4cm}

In the following sections, such steps are described in detail. For each of them, we discuss the challenges we had to overcome 
and the difficulties we had while conceiving the tools as asked by the projects' use-case partners. The methods we employed to address 
such challenges are presented together with the corresponding lessons learned.

\begin{figure}[h!]
	\centering
	\includegraphics[width=\linewidth]{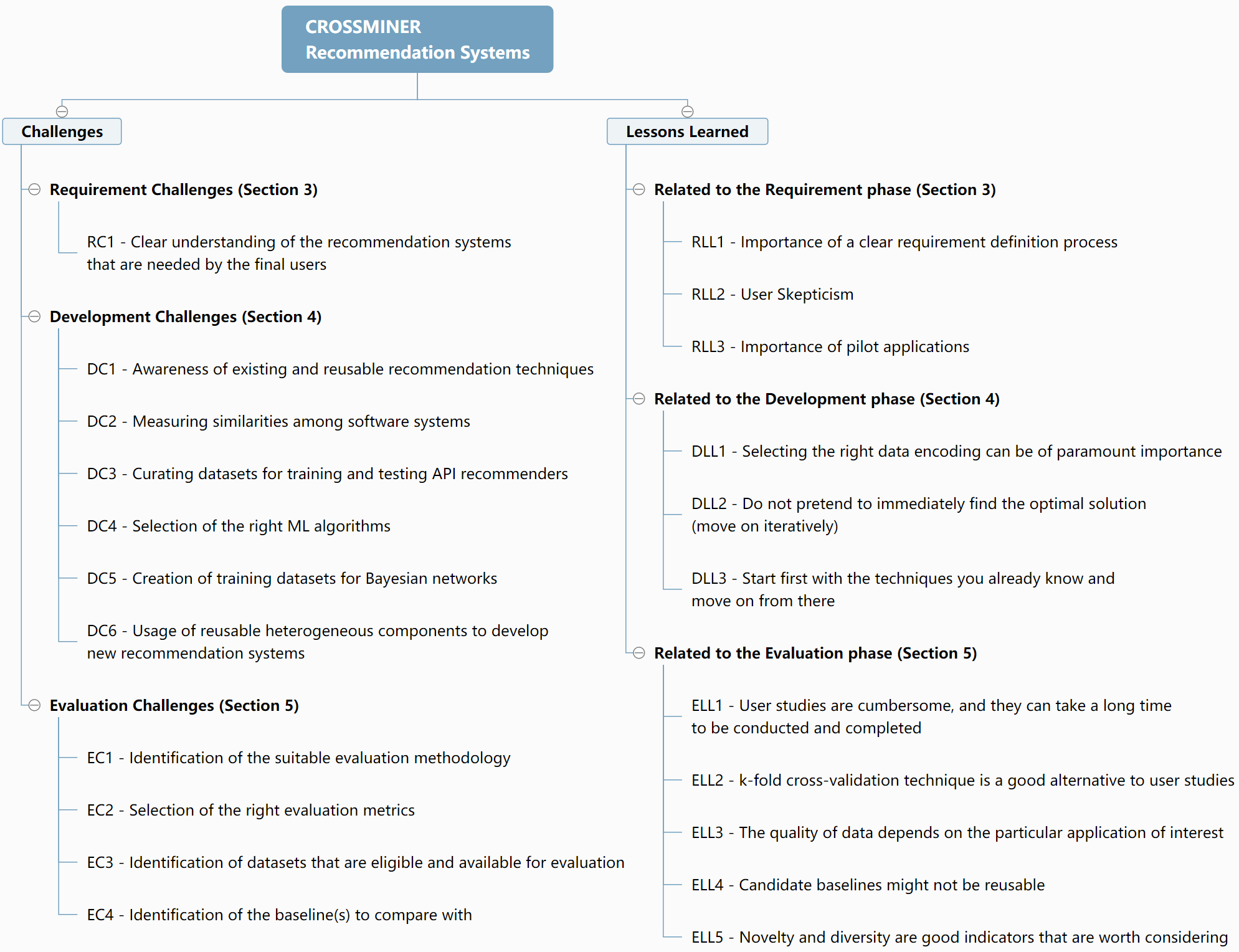}
	\caption{Map of challenges and lessons learned.}
	\label{fig:challengedandlessonslearnedmap}
\end{figure}

An overview of all the challenges and lessons learned are shown in the 
map depicted in 
Fig. \ref{fig:challengedandlessonslearnedmap}. For the sake of readability, 
challenges related to the requirement, development, and evaluation phases are 
identified with the three strings \emph{RC}, \emph{DC}, and \emph{EC}, respectively, followed by a cardinal number. %integer value. 
Similarly, the lessons learned are organized by 
distinguishing them with respect to the requirement (\textit{RLL}), development 
(\textit{DLL}), and evaluation (\textit{ELL}) phases.

%% file: src/requirements.tex
During the first six months of the project, we 
worked in tight 
collaboration with the industrial partners of 
the consortium to understand what 
they were expecting from the project 
technologies in terms of development 
support. For instance, the main business 
activity of one of the use-case 
partners consisted in the development and 
maintainance of a software quality 
assessment tool built atop of 
SonarQube.\footnote{\url{https://www.sonarqube.org/}}
 Whenever a 
new version of SonarQube was released, they 
needed to upgrade their tool to 
make 
it work with the new version of the SonarQube 
APIs. We, in turn, had to 
interact with 
the interested use-case partner to identify the 
types of recommendations that 
might have been useful for them to manage their 
API evolution problems. Other 
use case partners were asking for additional 
recommendations, which were posing 
requirements that might have had ripple effects 
on the other one's.

\subsection{Challenges}

\begin{figure}[t!]
	\centering
	\vspace{-.2cm}
	\includegraphics[width=0.9\linewidth]{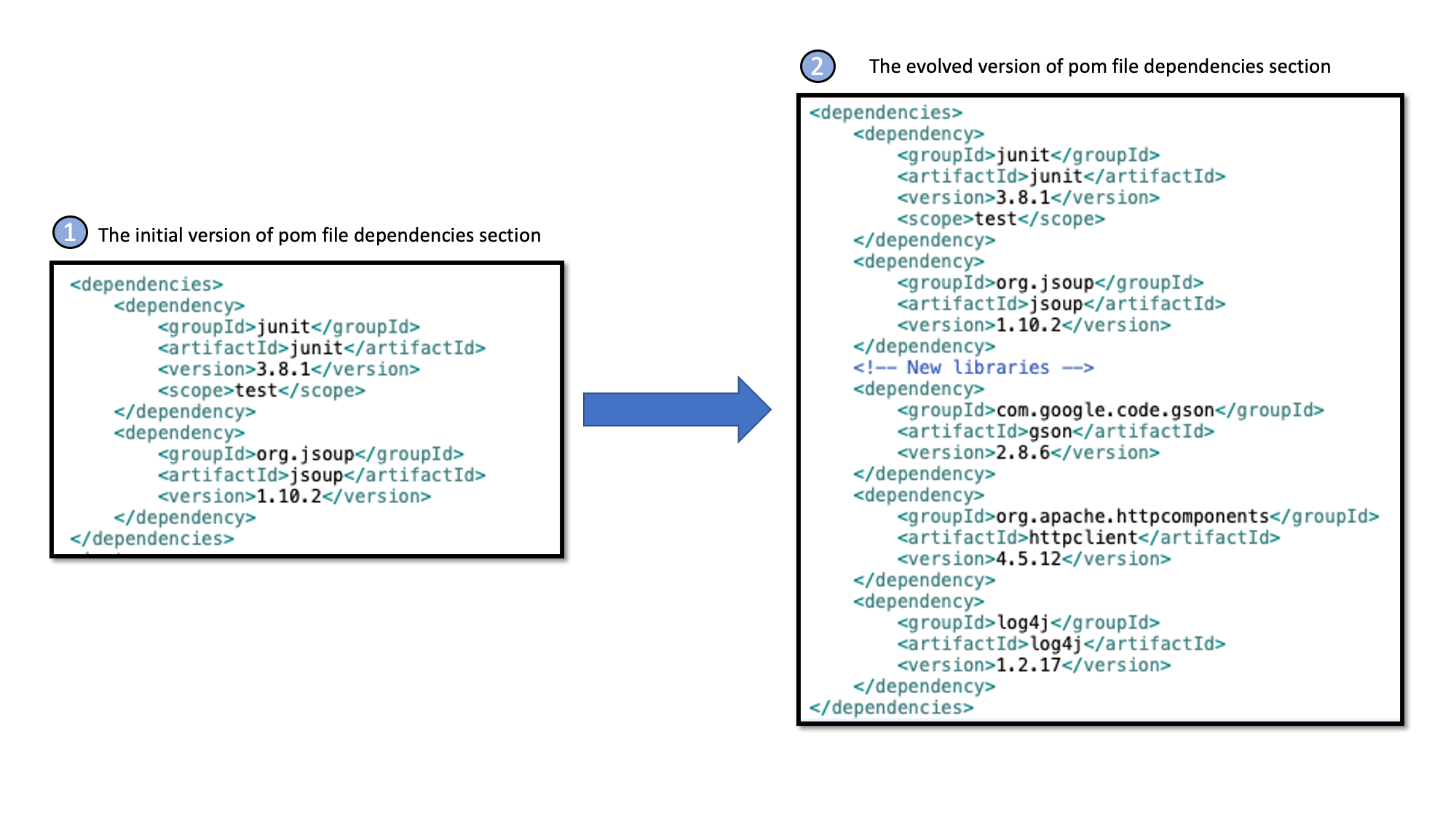}
	\caption{\code{pom.xml} files of a project 
	before (\emph{left}) and after 
	(\emph{right}) having adopted third-party 
	libraries recommended by \CR.}
	\vspace{-.2cm}
	\label{fig:CrossRec}	
\end{figure}

\challenge{RC1 - Clear understanding 
of the \rs that are needed by the 
end users}{%Understanding
Getting familiar with the 	
functionalities that are expected from the 
final users of the envisioned \rs is 
a daunting task. We might risk spending time 
on developing systems that are 
able to provide recommendations, which instead 
might not be relevant and in 
line with the actual user needs.}
\vspace{-.1cm}
\noindent
%
%conceived in the scope of CROSSMINER
%
% This is conformed with RC1, where we confirmed the importance of identifying the suitable techniques to build our recommendation 
%systems.
%
To deal with such a challenge and thus mitigate the risks of developing systems that might not be in line with the user requirements, 
we developed proof-of-concept \rs. In particular, we implemented demo projects that reflected real-world scenarios in terms of 
explanatory context inputs and corresponding recommendation items that the envisioned \rs should have produced. For instance, 
concerning \CR, we experimented on the \textit{jsoup-example}\footnote{\url{https://github.com/MDEGroup/FOCUS-user-evaluation}} 
explanatory Java project for scraping HTML 
pages. This project consists of source code and few related third-party libraries already included, \ie 
\emph{json-soup}\footnote{\url{https://jsoup.org/}} and \emph{junit}\footnote{\url{https://junit.org/}} as shown in the left-hand side 
of Fig.~\ref{fig:CrossRec}.
 
By considering such project as input, \CR 
provides a list of additional 
libraries as a suggestion that the project under 
development should also include. 
For instance, some beneficial libraries to be 
recommended are as follows: 
%\emph{i)} 
%\textit{commons-io}\footnote{\url{http://commons.apache.org/proper/commons-io/}}
% to assist with developing IO functionality; 
%\emph{ii)}  
%\textit{gson}\footnote{\url{https://github.com/google/gson}}
% for manipulating 
%JSON resources; \emph{iii)} 
%\textit{httpclient}\footnote{\url{https://hc.apache.org/}}
% for client-side 
%authentication, HTTP state management, and 
%HTTP connection management;  
%\emph{iv)} 
%\textit{log4j}\footnote{\url{https://logging.apache.org/}}
% to 
%enable 
%logging at runtime;  and \emph{v)} 
%\textit{commons-lang} that provides helper 
%utilities such as string manipulation methods, 
%object reflection, creation and 
%serialization. 
%\emph{i)} 
%\textit{commons-io}\footnote{\url{http://commons.apache.org/proper/commons-io/}}
% to assist with developing IO functionality; 
\emph{(i)}  
\textit{gson}\footnote{\url{https://github.com/google/gson}}
 for 
manipulating JSON resources; 
\emph{(ii)} 
\textit{httpclient}\footnote{\url{https://hc.apache.org/}}
 for 
client-side authentication, HTTP state 
management, and HTTP connection 
management; and \emph{(iii)} 
\textit{log4j}\footnote{\url{https://logging.apache.org/}}
 to enable logging at 
runtime.%;  and \emph{v)} \textit{commons-lang} 
%that provides helper utilities 
%such as string manipulation methods, object 
%reflection, creation and 
%serialization. 
%
%
%The possible recommended libraries are 
%discussed it is highly probable that 
By carefully examining the recommended 
libraries, we see that they have a 
positive impact on the project.
To be concrete, the usage of the 
\emph{httpcomponent} library allows the 
developer to access HTML resources by unloading 
the result state management and 
client-server authorization implementation on 
the library; meanwhile 
\emph{gson} could provide a parallel way to 
crawl public Web data; finally introducing a logging library,  \ie \emph{log4j}, can improve the project's maintainability.
%Candidates libraries should facilitate the 
%develops in getting the HTML page, 
%printing logging information, and manipulating 
%json objects.  
%Let consider the \textit{json-example} 
%demonstration 
%project\footnote{\url{https://github.com/MDEGroup/FOCUS-user-evalaution}}
% for 
%scraping html pages by using 
%json-soup\footnote{\url{https://jsoup.org/}} 
%external library. This is a maven project that 
%uses very only two external 
%libraries \ie 
%junit\footnote{\url{https://junit.org/}} and 
%json-soup. 

Concerning \FC, the process was a bit 
different, \ie use-case partners were 
providing us with incomplete source code 
implementation and their expectations 
regarding useful recommendations. Such 
artifacts were used as part of 
the requirements to implement the system able 
to resemble them. The use-case 
partner expects to get code snippets that 
include suggestions to improve the 
code, and predictions on next API function 
calls.

%For this purpose
To agree with the use-case partners on the 
recommendations that were expected 
from \FC, we experimented on a partially 
implemented method of the 
\textit{jsoup-example} project named 
\code{getScoresFromLivescore} shown in 
Listing~\ref{lst:partial-implemented-method}. 
The method should be designed so 
as being able to collect the football scores 
listed in the \url{livescore.com} 
home page. To this end, a JSON document is 
initialized with a connection to the 
site URL in the first line. By using the JSOUP 
facilities, the list of HTML 
element  of the class \code{sco} is stored in 
the variable \code{score} in the 
second line. Finally, the third line updates 
the scores with all of the parents 
and ancestors of the selected scores elements.
%From possible recommendation the first one 
%suggests the usage of 
%\code{userAgent} method to avoid that some 
%sites block the http requests.
%The blue box consists of the recommendation 
%for improving the code (\ie the 
%usage of \code{userAgent} method to avoid that 
%some sites block the http 
%requests), and  to predict the next 
%\emph{jsoup} invocation.
%Furthermore, some recommendation can include 
%API function call to a competitor 
%library 
%\emph{HTMLUnit}\footnote{\url{http://htmlunit.sourceforge.net/}},
% and 
%others that belong to the \emph{jsoupcrawler} 
%extension. It is worth noting 
%that \emph{HTMLUnit} is a direct competitor of 
%\emph{jsoup} that includes 
%different browser user agent implementation, 
%and \emph{jsoupcrawler} is a 
%custom extension of \emph{jsoup}. Moreover, 
%the first recommended api function 
%call 

\begin{lstlisting}[language=Java,caption=Partial
 implementation of the 
explanatory \code{getScoresFromLivescore()} 
method., 
label=lst:partial-implemented-method]
public static void getScoresFromLivescore() 
throws IOException {
  Document document = 
  Jsoup.connect("https://www.livescore.com/").get();
  Elements scores = 
  document.getElementsByClass("sco");
  scores = scores.parents();
  ...
}
\end{lstlisting}

\begin{figure}[t!]
	\centering
	\includegraphics[width=0.68\linewidth]{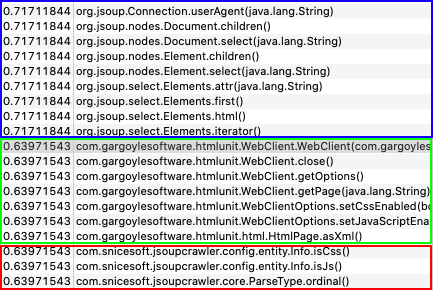}
	\caption{Recommended API calls for the 
	\code{getScoresFromLivescore()} 
	method in Listing~\ref{lst:partial-implemented-method}.}
	\label{fig:focusinaction}
	\vspace{-.5cm}
\end{figure}

Figure~\ref{fig:focusinaction} depicts few 
recommendations that our use-case 
partners expected when we presented the example 
shown in Listing 
\ref{lst:partial-implemented-method}. The blue 
box contains the recommendation 
for improving the code, \ie the 
\code{userAgent} method is to prevent sites 
from blocking HTTP requests, and to predict the 
next \emph{jsoup} invocation.
Furthermore, some recommendations could be 
related to API function calls of a 
competitor library or extension.
For this reason, the green and red boxes 
contain invocations of 
\emph{HTMLUnit},\footnote{\url{http://htmlunit.sourceforge.net/}} a direct 
competitor of \emph{jsoup} that includes 
different browser user agent 
implementations, and \emph{jsoupcrawler} a 
custom extension of \emph{jsoup}. %, 
\FC has been conceptualized to suggest to developers recommendations consisting of a list of API method calls that should be used next. Furthermore, it also recommends real code snippets that can be used as a reference to support developers in finalizing the method definition under development. More code examples provided by \FC are available in an online appendix.\footnote{\url{https://github.com/crossminer/FOCUS}}

%respectively.
%is . Moreover, the first recommended API 
%function call 

%\revised{More examples of \FC recommendations are provided in an online appendix.\footnote{\url{https://mdegroup.github.io/FOCUS-Appendix/}}}

%API function call to a competitor library 
%\emph{HTMLUnit}\footnote{\url{http://htmlunit.sourceforge.net/}}

\subsection{Lessons learned} % from defining the requirements of the \projectName \rs

\begin{figure}[t!]
	\centering
	\includegraphics[width=\linewidth]{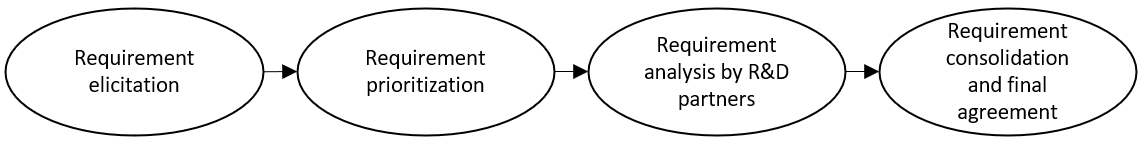}
	\caption{Requirement definition process.}
	\label{fig:requirementprocess}
\end{figure}

\textbf{RLL1 -- Importance of a clear 
requirement definition process:} 
As previously mentioned, we managed to address 
Challenge RC1 through a 
tight collaboration with the use case partners. 
In particular, we applied the 
requirement definition process shown in Fig.~\ref{fig:requirementprocess}, 
which consists of the following steps and that 
in our opinion can be applied 
even in contexts that are different from the \projectName one:

\begin{itemize}

	\item[--] \textit{Requirement 
	elicitation:} The final user 
	identifies use cases that are 
	representative and that identify the 
	functionalities that the wanted \rs should 
	implement. By considering such 
	use cases, a list of requirements is 
	produced;
	
	\item[--] \textit{Requirement 
	prioritization:} The list of 
	requirements produced in the previous step 
	can be very long, because users 
	tend to add all the wanted and ideal 
	functionalities even those that might 
	be less crucial and important for them. For 
	this reason, it can be useful
	to give a priority to each requirement in 
	terms of the modalities 
	\textit{shall}, \textit{should}, and 
	\textit{may}. \textit{Shall} is used 
	to denote essential requirements, which are 
	of highest priority for 
	validation of the wanted \rs. 
	\textit{Should} is used to denote a 
	requirement that would be not essential 
	even though would make the wanted 
	\rs working better. \textit{May} is used to 
	denote requirements that would 
	be 	interesting to satisfy and explore even 
	though irrelevant for 
	validating the wanted technologies;
	
	\item[--] \textit{Requirement 
	analysis by R\&D partners:} The 
	prioritized list of requirements is 
	analyzed by the \textit{research and 
	development} partners with the aim of 
	identifying the major components that 
	need to be developed. Possible 
	technological challenges that might 
	compromise the satisfaction of some 
	requirements are identified in this 
	step and considered in the next step of the 
	process;
		
	\item[--] \textit{Requirement 
	consolidation and final agreement:} 
	By considering the results of the analysis 
	done by the R\&D partners, the 
	list of requirements is further refined and 
	consolidated. After this step, 
	user case partners have ensured highest 
	priority requirements, which will 
	be implemented by R\&D partners.
	
\end{itemize} 

We have applied such a process in 
different projects and we 
successfully applied it also for developing the 
\rs that we identified in the context of the \projectName project.

\smallskip
\noindent
\textbf{RLL2 -- Users skepticism:} 
Especially at the early stages of the wanted \rs development, target 
users might be skeptical about the relevance of the potential items that can be recommended. We believe that defining the requirements 
of the desired \rs in tight collaboration with the final users is the right way to go. Even when the proposed approach has been 
evaluated employing adequate metrics, final users still might not be convinced about the retrievable recommendations' relevance. User 
studies can be one of the possible options to increase the final users' trust, even though a certain level of skepticism might remain 
when the intended final users have not been involved in the related user studies.

\smallskip
\noindent
\textbf{RLL3 -- Importance of pilot 
applications:} Using a pilot application can be beneficial to support the interactions between the final users and the developers of 
the wanted \rs. The application can allow the involved parties to tailor the desired functionalities utilizing explanatory inputs and 
corresponding recommendations that the envisioned system should produce.

%\nb{Davide}{@Juri we need to mention some 
%concrete example here.}

%\nb{Davide}{Explain how we coped with such a 
%challenge.}

%% file: src/Developing.tex
Once the requirements of the expected RSs were 
agreed with the use-case partners, we started with the development of each 
identified RS. 

\smallskip
\challenge{DC1 -- Awareness of existing and reusable recommendation 
techniques}{Over the last decades, several \rs have been developed by both 
academia and 
	industry. When realizing a new type of recommendation system, it is crucial 
	to have a 
	clear knowledge of the possible techniques and patterns that might be 
	employed to develop new ones. Since the solution space is extensive, 
	comparing and 
	evaluating candidate approaches can be a very daunting task.}
\vspace{-.1cm}
\noindent
To overcome such challenge we performed a rigorous literature review by reviewing related studies emerging from premier venues 
in the Software Engineering domain, \ie conferences such as ICSE, ASE, SANER, or journals such as TSE, TOSEM, JSS, to name a few. Being aware of existing systems is also important for the evaluation phase. For studying a recommendation system, besides conventional quantitative and qualitative evaluations, it is necessary to compare it with state-of-the-art approaches. Such an 
	issue is also critical in other domains, \eg Linked Data~\cite{DBLP:conf/rweb/NoiaO15}, or music recommendations \cite{10.1145/2740908.2742141,DBLP:journals/ijmir/SchedlZCDE18}. When we started with the design of the systems, the solution space was huge, considering the use-case partners’ requirements. However, being aware of what already exists is very important to save 
	time, resources, and avoid the reimplementation of already existing techniques and tools.
%
%
%Though that was 
%also a strenuous activity, 
To this end, by analyzing the existing 
literature about recommendation systems we identified and modeled their
relevant variabilities and commonalities. 

Section \ref{sec:designFeatures} brings in the main design features of 
our \rs. 
Specific challenges that have faced to design the different tools are described 
in Sections 
\ref{sec:CsCrChallenges}--\ref{sec:MndnChallenges}. The lessons learned by designing the \projectName \rs are discussed in Section~\ref{sec:designLessons}.

\subsection{Main design features}\label{sec:designFeatures}

\begin{figure}[t!]
	\centering
	\includegraphics[width=\linewidth]{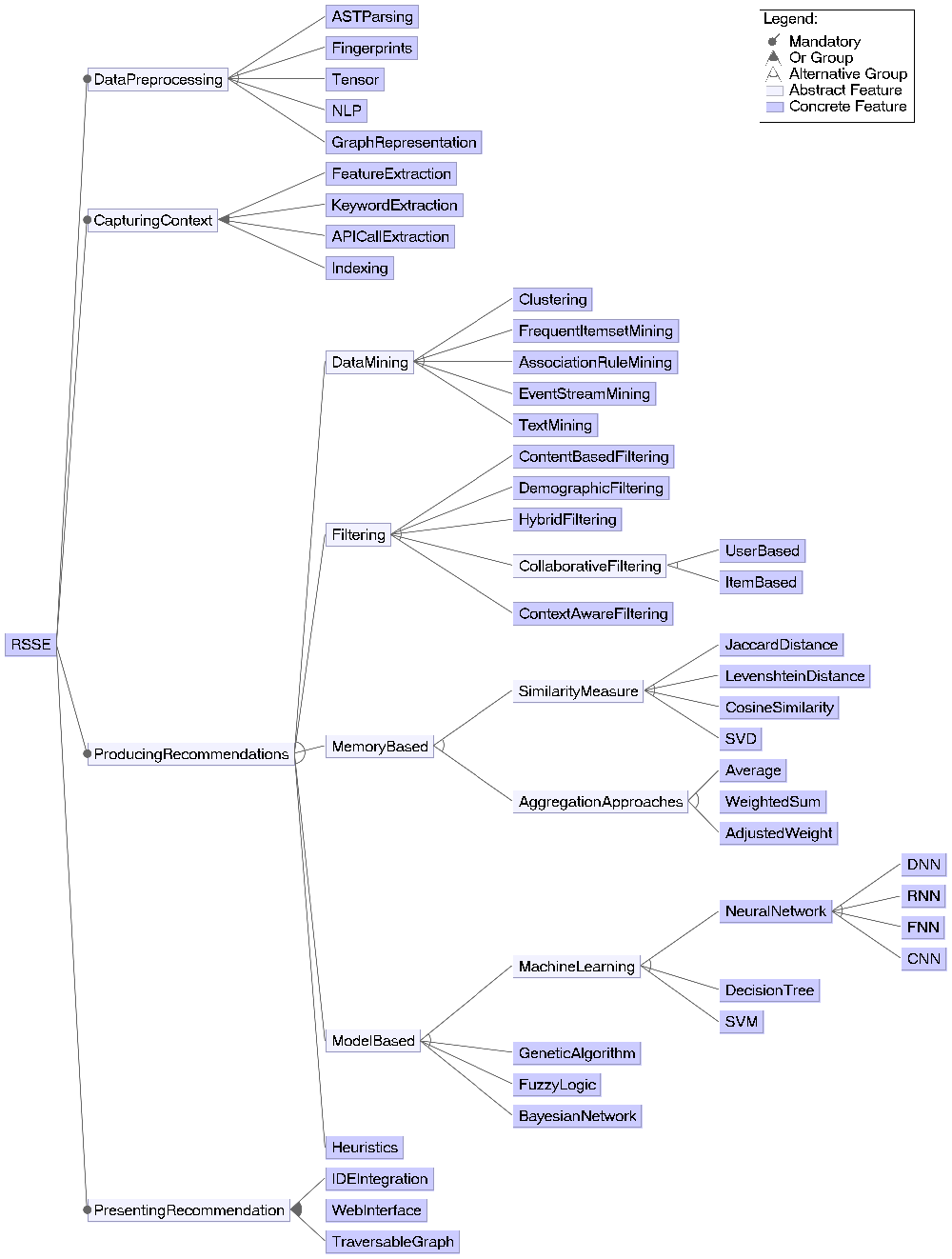}
	\caption{Main design features of recommendation systems in software 
	engineering.}%[Increase font size, reorder features according to the text flow]
	\label{fig:RecsysFeatures}
\end{figure}

Our results are documented using
feature diagrams which are a common notation in 
domain analysis \cite{Czarnecki02}. Figure \ref{fig:RecsysFeatures}
shows the top-level features or \rs, \ie \emph{Data Preprocessing},  
\emph{Capturing Context}, \emph{Producing Recommendations}, and 
\emph{Presenting Recommendations} in line with the main functionalities 
typically implemented by \rs.
We extracted all the shown components mainly from existing studies 
\cite{Bobadilla2013109,laser_software_2015,robillard_recommendation_2014}
 as well as from our development 
experience under the needs of the \projectName project.  The top-level features 
shown in Fig. \ref{fig:RecsysFeatures} are described below.

\medskip
\noindent
\textit{Data Preprocessing:} In this phase techniques and tools are applied to 
extract valuable information from different data sources according to their 
nature.  In particular, \textit{structured data} adheres to several rules that 
organize elements in a well-defined manner. Source code and XML documents are 
examples of this category. Contrariwise, \textit{unstructured data} may 
represent different content without defining a methodology to access the data. 
Documentation, blog, and plain text fall into this category. Thus, the data 
preprocessing component must be carefully chosen considering the features of 
these miscellaneous sources. 

\code{ASTParsing} involves the analysis of structured data, typically the 
source code of a given software project. Several libraries and tools are 
available to properly perform operations on ASTs, \eg fetching function calls, 
retrieving the employed variables, and analyzing the source code dependencies. 
Additionally, snippets of code can be analyzed using \code{Fingerprints}, \ie a 
technique that maps every string to a unique sequence of bits. Such a strategy 
is useful to uniquely identify the input data and compute several operations on 
it, \ie detect code plagiarism as shown in \cite{DBLP:conf/aics/ZhengPL18}.

Moving to unstructured input, \code{Tensors} can encode mutual relationships 
among data, typically users' preferences. Such a feature is commonly exploited 
by collaborative filtering approaches as well as by heavy computation on the input data to produce recommendations. Plain text is the most 
spread type of unstructured data and it includes heterogeneous content, \ie API 
documentation, repository's description, Q\&A posts, to mention a few. A real 
challenge is to extract valuable elements without losing any relevant 
information. Natural processing language (\code{NLP}) techniques are employed 
to perform this task by means of both syntactic and semantic analysis. 
Stemming, lemmatization, and tokenization are the main strategies successfully 
applied in existing \rs. Even the \textit{\bnetwork}~approach 
previously presented employs such techniques as preparatory task before 
 the training phase.
Similarly to tensors, \code{GraphRepresentation} is useful to model reciprocal 
associations among considered elements. Furthermore, graph-based data encodings 
can be used to find peculiar patterns considering nodes and edges semantic. 

\medskip
\noindent
\textit{Capturing Context:} After the data preprocessing phase, the developer 
context is excerpted from the programming environment to enable the 
underpinning recommendation engine. A well-founded technique primarily employed 
in the ML domain is the \texttt{FeatureExtraction} to concisely represent the 
developer's context. Principal Component Analysis (PCA) and Latent Semantic 
analysis (LDA) are just two of such techniques employed for such a purpose. 
\code{Keyword extraction} and \code{APICallExtraction} are two techniques 
mostly used when the \emph{Capturing Context} phase has to analyze source code. 
Capturing context often involves the search over big software projects. Thus, a 
way to store and access a large amount of data is necessary to speed up the 
recommendation item delivery. \code{Indexing} is a technique mainly used by the 
code search engines to retrieve relevant elements in a short time. 
%In particular, this technique allows storing indexes in a database and query 
%it.

\medskip
\noindent
\textit{Producing Recommendations:} In this phase, the actual recommendation 
algorithms are chosen and executed to produce suggestions that are relevant for 
the user context, once it is previously captured. By several variating 
parameters such as type of the required input and the underlying structure, we 
can elicit different features as represented in the diagram shown in Fig. 
\ref{fig:RecsysFeatures}. Concerning \code{Data Mining} techniques, some of 
them are based on pattern detection algorithms, \ie~\code{Clustering}, 
\code{FrequentItemsetMining}, and \code{AssociationRule\-Mining}. 
\texttt{Clustering} is usually 
applied %in the classification problem 
to group objects according to some 
similarity functions. The most common algorithm is the K-means based on 
minimizing the distance among the items. A most representative element called 
centroid is calculated through a linkage function. After such a computation, 
this algorithm can represent a group of elements by referring to the most 
representative value. \texttt{FrequentItemsetMining} aims to group items with 
the same frequencies, whereas \texttt{AssociationRuleMining} uses a set of 
rules to discover possible semantic relationships among the analysed elements. 
%The apriori algorithm can support both strategies by first identifying the 
%itemsets and then inferring the association rules.
%
Similarly, the \texttt{EventStreamMining} technique aims to find recurrent patterns in data streams. 
A stream is defined as a sequence of events usually represented by a Markov chain. 
Through this model, the algorithm can exploit the probability of each event to establish relationships 
and predict a specific pattern. Finally, \texttt{TextMining} techniques often involve information 
retrieval concepts such as entropy, latent semantic analysis (LSA), or extended boolean model.
 In the context of producing recommendations, such strategies can be used to find similar terms by 
 exploiting  different probabilistic models that analyze the correlation among textual documents. 

The availability of users' preferences can affect the choice of recommendation algorithms. \code{Filtering} strategies 
dramatically exploit the user data, \eg their ratings assigned to purchased 
products. \code{Content\-BasedFiltering} (CBF) employs historical data 
referring 
to items with positive ratings. It is based on the assumption that items with 
similar features have the same score. Enabling this kind of filtering requires 
the extraction of the item attributes as the initial step. Then, CBF compares 
the set of active items, namely the context, with possible similar items using 
a similarity function to detect the closer ones to the user's needs. 
\code{DemographicFiltering} compares attributes coming from the users 
themselves instead of the purchased items. These two techniques can be combined 
in \code{HybridFiltering} techinques to achieve better results. 

So far, we have analyzed filtering techniques that tackle the features of items and users. \code{CollaborativeFiltering} (CF) approaches analyze the 
user's behaviour directly through its interaction with the system, \ie the 
rating activity. \code{UserBased} CF relies on explicit feedback coming from 
the users even though this approach suffers from scalability issues in case of 
extensive data. The \code{ItemBased} CF technique solves this issue by 
exploiting users' ratings to compute the item similarity. Finally, 
\code{ContextAwareFiltering} involves information coming from the environment, \ie temperature, geolocalization, and time, to name a few. Though this kind 
of filtering goes beyond the software engineering domain, we list it to 
complete the filtering approaches landscape.

The \code{MemoryBased} approach acts typically on user-item matrixes to compute 
their distance involving two different methodologies, \ie~\code{Similarity\-Measure} and \code{AggregatationApproach}. The former 
involves 
the evaluation of the matrix similarity using various concepts of similarity. 
For instance, \code{JaccardDistance} measures the similarity of two sets of 
items based on common elements, whereas the \code{LevenshteinDistance} is based 
on the edit distance between two strings. Similarly, the 
\code{CosineSimilarity} measures the euclidean distance between two elements.  
Besides the concept of similarity, techniques based on matrix factorization are employed to make the 
recommendation engine more scalable. 
Singular value decomposition (\code{SVD}) is a technique being able to reduce the dimension of 
the matrix and summarize its features.
Such a strategy is used to cope with a large amount of data, even though it is computationally expensive.
\code{Aggregatation\-Approach}es analyze relevant statistical information of 
the 
dataset such as the variance, mean, and the least square. To mitigate bias lead 
by the noise in the data, the computation of such indexes use adjusted 
weigths as a coefficient to rescale the results. 

To produce the expected outcomes, \code{MemoryBased} approaches require the 
direct usage of the input data that cannot be available under certain 
circumstances. Thus, \code{ModelBased} strategies can overcome this limit by 
generating a model from the data itself. \code{MachineLearning} offers several 
models that can support the recommendation activity. \code{NeuralNetwork} models can 
learn a set of features and recognize items after a training phase. By 
exploiting different layers of neurons, the input elements are labeled with 
different weights. Such values are recomputed during different training rounds 
in which the model learns how to classify each element according to a 
predefined loss function. Depending on the number of layers, the internal 
structure of the network, and other parameters, it is possible to use different 
kinds of neural networks including Deep Neural Networks  (\code{DNN}), 
Recurrent Neural Networks (\code{RNN}), Feed-forward Neural Networks 
(\code{FNN}), or Convolutional Neural Networks (\code{CNN}). 
Besides ML models, a recommendation system can employ several models to suggest relevant items. 
\code{GeneticAlgorithm}s are based on evolutionary principles that hold in the 
biology domain, \ie natural species selection. \code{FuzzyLogic} relies on a logic 
model that extends classical boolean operators using continuous variables. In 
this way, this model can represent the real situation accurately. Several 
probabilistic models can be used in a recommendation system. \code{BayesianNetwork} is mostly employed to classify unlabeled data, although 
it is possible to employ it in recommendation activities.

Besides all these well-founded techniques, recommended items can be produced 
by means of \texttt{Heuristics} techniques to encode the knowhow of domain experts. 
Heuristics employ different approaches and techniques together to obtain better 
results as well as to overcome the limitations of other techniques. On the one 
hand, heuristics are easy to implement as they do not rely on a complex 
structure. On the other hand, they may produce results that are sub-optimal 
compared to more sophisticated techniques. 
%The Strathcona tool \cite{holmes_strathcona_nodate} is an example in which 
%such techniques are successfully employed.

\medskip
\noindent
\textit{Presenting Recommendations:} As the last phase, the produced 
recommendation items need to be properly presented to the developer. To this 
end, several strategies involve potentially different technologies, including 
the development of extensions for IDEs and dedicated Web-based interfaces.
\texttt{IDEIntegration} offers several advantages, \ie 
auto-complete shortcuts and dedicated views showing the recommended items. The 
integration is usually performed by the development of a 
plug-in, as shown in existing \rs 
\cite{DBLP:conf/kbse/LvZLWZZ15,DBLP:journals/ese/PonzanelliBPOL16}.
Nevertheless, developing such an artifact requires much effort, and the 
integration must take into account possible incompatibilities among all 
deployed components. A more flexible solution is represented by 
\texttt{WebInterface}s in which the recommendation system can be used as a 
stand-alone platform. Even though the setup phase is more accessible rather than the IDE 
solution, presenting recommendations through a web service must handle some 
issues, including server connections, and suitable response times. 
%Except for the MNBN, we develop and integrate all the presented tool in the 
%Scava platform that combines these two presentation techniques.
For presentation purposes, interactive data structures might be useful in navigating the recommended 
items.
 \texttt{TraversableGraph} is just one successful example of this category. Strathcona~\cite{holmes_strathcona_nodate} makes use of this technique to show the snippets 
of code rather than simply retrieving them as ranked lists. In this way, final 
users can figure out additional details about the recommended items.
%, \ie required inputs and produced outputs.

%For each dimension represented as boxes in Fig. \ref{fig:featuremodel}, 
%corresponding features (and sub-features) are depicted as ovals to represent 
%possible techniques that can be employed. Dashed boxes depict attributes of 
%the 
%mandatory features to include more information about them. For instance, data 
%preprocessing can be employed on source code, Q\&A items like StackOverflow 
%posts, \etc. whereas data mining and filtering techniques are some of the 
%approaches that can be employed to produce recommendations.

%In this section we describe the process that has been conceived and applied to evaluate the performance of \CS compared to some baselines. W

%by considering the analogy of typical applications of RDF graphs and the 
%problem of detecting the similarity of open source projects, we developed 
%\CrossSim ({\bf C}ross project {\bf R}elationships for computing {\bf O}pen 
%{\bf S}ource {\bf S}oftware {\bf Sim}ilarity)~\cite{8498236}, an approach that 
%makes use of graphs for modeling different types of relationships in the OSS 
%ecosystem \cite{DBLP:conf/iir/NguyenRR18}. Similar to RDF graphs, the 
%representation model can 
%In this sense, choosing the right technique to compute software/projects similarity is a question that may arise at any time. 

\subsection{Development challenges for \CS and 
\CR}\label{sec:CsCrChallenges}

%\todo{Describe the main difficulties here: first of all which underpinning 
%algorithm? Why we adopted a graph based representation? The choice has been 
%done by considering the fact that some of the authors already had expertice in 
%graph similarity algorithms and we followed the intuition that the know 
%techniques might have been employed if the projects are properly represented. 
%We started with a set of very few projects with the aim of setting up the 
%context needed for enabling the technique to a bigger dataset.}

\challenge{DC2 -- Measuring similarities among 
	software systems}{Considering
	 the miscellaneousness of artifacts in open source software repositories, 
	similarity computation becomes more complicated as many artifacts and 
	several cross relationships prevail.}
\vspace{-.1cm}
\noindent
In OSS forges like \GH, there are several connections and 
interactions, such as development commit to repositories, user star 
repositories, or projects contain source code files, to mention a few. To 
conceptualize \CS~\cite{Nguyen:2019:JSS:CrossSim}, we came up with the 
application of a 
graph-based representation to capture the semantic features among various actors, and consider their 
intrinsic connections. We modeled the community of 
developers together with OSS projects, libraries, source code, and their mutual 
interactions as an \emph{ecosystem}. In this system, either humans or non-human 
factors have mutual dependencies and implications on the others. Graphs allow 
for flexible data integration and facilitates numerous similarity metrics 
\cite{Blondel:2004:MSG:1035533.1035557}. 

We decided to adopt a graph-based representation to deal with the 
project similarity issue because some of the co-authors already addressed a 
similar problem in the context of Linked Data. The analogy of the two problems inspired us to apply %make an attempt of applying 
the similarity technique already developed \cite{10.1145/2740908.2742141} to 
calculate the similarity of representative software projects. The initial 
evaluations were encouraging and consequently, we moved on by refining the 
approach and improving its accuracy.

Despite the need to better support software developers while they are programming, very few works have been conducted concerning the 
techniques that facilitate the search for suitable third-party libraries from OSS repositories. 
We designed and implemented \CR on top of \CS: the graph representation was 
exploited again to compute similarity among software projects, and to provide 
inputs for the recommendation engine. 

Understanding the features that are relevant for 
the similarity calculation was a critical task, which required many iterations 
and evaluations. For instance, at the beginning of the work we were including 
in the graph encoding information about developers, source code, \GH star 
events when available, etc. However, by means of the performed experiments, we 
discovered that encoding only dependencies and star events is enough 
to get the best performance of the similarity approach 
\cite{Nguyen:2019:JSS:CrossSim}.

To sum up, concerning the features shown in Fig. 
\ref{fig:RecsysFeatures}, both \CS and \CR make use of a graph-based 
representation for supporting the \textit{Data Preprocessing} activity. 
Concerning the \textit{Producing Recommendation} phase, item-based 
collaborative filtering techniques have been exploited. For the 
\textit{Capturing Context} phase, the project being developed is encoded in terms of different features, including used third-party 
libraries, and README 
files. Recommendations are presented to the user directly in the used
Eclipse-based development environment.

%Among the existing approaches, to the best of our knowledge, LibRec~\cite{6671293},~LibFinder~\cite{Ouni:2017:SSL:3032135.3032325}, and~LibCUP \cite{SAIED2018164} are the most advanced techniques for library recommendation to support OSS developers. The systems have been demonstrated to be able to recommend project libraries with a high \emph{success rate}. With \CR, we inherited the graph representation proposed to. 

%\nb{Davide}{@Phuong @Juri any specific challenge related to \CS?}

%Concerning \FC 
%Over the past decade, the problem of API learning has garnered considerable interest from the research community.

\subsection{Development challenges for \FC}

During the development process, rather than programming from scratch, 
developers look for libraries that implement the desired functionalities and 
integrate them into their existing projects \cite{DBLP:conf/iir/NguyenRR18}. 
For such libraries, API function calls are the entry point which allows one to 
invoke the offered functionalities. However, in order to exploit a library to 
implement the required feature, programmers need to consult various sources, 
\eg API documentation to see how a specific API instance is utilized in the 
field. Nevertheless, from these external sources, there are only texts 
providing generic syntax or simple usage of the API, which may be less relevant 
to the current development context. In this sense, concrete examples of source 
code snippets that indicate how specific API function calls are deployed in 
actual usage, are of great use \cite{Moreno:2015:IUT:2818754.2818860}.

Several techniques have been developed to 
automate the extraction of API usage 
patterns~\cite{Robillard:2013:AAP:2498733.2498776} in order to reduce 
developers' burden when manually searching these sources and to provide them 
with high-quality code examples. However, these techniques, based on 
clustering~\cite{niuAPIUsagePattern2017,Wang2013Mining,10.1007/978-3-642-03013-0_15}
 or predictive modeling~\cite{Fowkes:2016:PPA:2950290.2950319}, still suffer 
from high redundancy and poor run-time 
performance. 

By referring to the features shown in Fig. 
\ref{fig:RecsysFeatures}, differently from other existing approaches which 
normally rely on 
clustering to find API calls, \FC implements a \textit{context-aware 
collaborative-filtering} 
system that exploits the cross relationships among different artifacts in OSS 
projects to represent them in a graph and eventually to predict the inclusion 
of additional API invocations. Given an active declaration representing the 
user \textit{context}, we search for prospective invocations from those in 
similar declarations belonging to comparable projects. Such a phase is made 
possible by a proper \textit{data preprocessing} technique, which encodes the 
input data by means of a \textit{tensor}. The main advantage 
of our tool is that it can recommend real code snippets that match well with the 
development context. In contrast with several existing approaches, 
\FC does not depend on any specific set of libraries and just needs OSS 
projects as background data to generate API function calls. More importantly, the system scales well with large datasets
using the collaborative-filtering technique that filters out irrelevant items, 
thus improving efficiency. The produced recommendations are 
shown to the users directly in the Eclipse-based IDE.

\smallskip
\challenge{DC3 - Curating a dataset for training and testing API 
recommenders requires significant effort}{To provide input data in 
the form of a tensor, it is necessary to parse projects for extracting their 
constituent declarations and invocations.}
\vspace{-.2cm}
\noindent
A major obstacle that we needed to overcome when implementing \FC is as follows. To provide input data in the form of a tensor, it was 
necessary to 
parse projects to extract their constituent declarations and invocations. 
However, \FC relies on Rascal \cite{7070485} to function, which in turn works only with compilable Java source code. To this end, we 
populated training data for the system from two independent sources. First, we curated a set of Maven jar files which were compilable 
by their nature. Second, we crawled and filtered out to select only GitHub projects that contain informative 
\emph{.classpath}, which is an essential requirement for running Rascal. Once the tensor has been properly formulated, \FC can work on 
the collected background data, being independent of its origin. One of the considered datasets was 
initially consisting of 5,147 Java projects retrieved from the Software 
Heritage archive \cite{SH}. To satisfy the baseline constraints, we first restricted the dataset to the list of projects that use at 
least one of the considered third-party libraries. Then, to comply with the requirements of 
\FC, we restricted the dataset to those projects containing at least one 
\code{pom.xml} file. Because of such constraints, we ended up with a 
dataset 
consisting of 610 Java projects. Thus, we had to create a dataset ten times bigger than the used one for the evaluation.

%We cannot use the ShareJar dataset from D6.4 to evaluate FOCUS 
%since the dataset does not contain enough information that can be used to run Rascal to parse the original data into the M3 format, which is an essential requirement for running FOCUS. The set of 3,600 Maven projects has been selected as the projects exhibit all the necessary criteria to be provided as input for Rascal, and thus they are suitable for evaluating FOCUS.

%\nb{Davide}{@Phuong @Juri any specific challenge related to the development of 
%\focus? Is there any specific challenge we can highlight here? For instance, 
%for \CS and \CR above we mentioned the issue of identifying the best 
%configuration.}

%Concerning \bnetwork ... %\nb{Davide}{@Claudio @Riccardo any specific challenge related to \CS?}

\subsection{Development challenges of \bnetwork} 
\label{sec:MndnChallenges}

In recent years, \GH has been at the forefront of platforms for 
storing, 
	analyzing and maintaining the community of OSS projects. To foster the 
	popularity and reachability of their repositories, \GH users make daily 
	usage of the star voting system as well as forking 
	\cite{borges_whats_2018,Jiang:2017:WDF:3042021.3042043}. These 
	features allow for increasing the popularity of a certain project, even 
	though the 
	search phase has to cope with a huge number of items. To simplify this 
	task, 
	\GH has introduced in 2017 the concept of \emph{topics}, a list of tags 
	aiming to describe a project in a succinct way. Immediately after the 
	availability of the topics concept, the 
	platform introduced Repo-Topix 
	\cite{ganesan_topic_2017} to assist developers when creating new projects 
	and 
	thus, they have to identify representative topics. 
	%As discussed in \cite{10.1145/3383219.3383227}, 
	Though Repo-Topix is 
	already in place, there are rooms for improvements, \eg in terms of the 
	coverage of the recommended topics, and of the underpinning analysis 
	techniques. To this end, we proposed MNBN~\cite{10.1145/3383219.3383227}, an approach based on a Multinomial Naive Bayesian network technique to automatically recommend topics given the \RM 
	file(s) of an input repository. 

The main challenges related to the development of \bnetwork ~concern three main 
dimensions as follows: \textit{(i) identification of the underpinning algorithm}, 
\textit{(ii) creation of the training dataset}, and \textit{(iii) usage of 
heterogeneous reusable complements}, and they are described below.

\smallskip
\noindent
\challenge{DC4 -- Selection of the right ML algorithms}{Due 
to the well-known no-free lunch theorem that holds for any Machine Learning 
(ML) approach 	\cite{wolpert_no_1997}, selecting the suitable model for the 
problem at hand is one of the possible pitfalls.}

\vspace{-.1cm}
\noindent
Concerning the Machine Learning domain, all relevant results have been obtained through empirical observations undertaken on different 
assessments. Thus, to better 
understand the context of the addressed problem we analyzed existing 
approaches that deal with the issue of text classification using ML models. 
Among the analyzed tools, the Source Code Classifier (SCC) tool 
\cite{Alreshedy2018SCCAC} can classify code snippets using the MNB network as the underlying model. In particular, this tool discovers 
the programming language of each snippet coming from StackOverflow posts. The results show that Bayesian networks outperform other 
models in the textual analysis task by obtaining 75\% of accuracy and success rate. 
Furthermore, there is a subtle correlation between the Bayesian classifier and the TF-IDF weighting 
scheme~\cite{hutchison_multinomial_2004}. A comprehensive study has been 
conducted by comparing TF-IDF with the Supporting Vector Machine (SVM) 
using different datasets. The study varies the MNB 
parameters to investigate the impacts of the two mentioned preprocessing techniques for each of them. The evaluation demonstrates that 
the TF-IDF scheme leads to better prediction performance than the SVM technique. Thus, we decided to adopt the mentioned MNBN 
configuration considering these two findings \textit{(i)} this model can adequately classify text content and \textit{(ii)} the TF-IDF 
encoding leads benefits in terms of overall accuracy.

\smallskip
\noindent
\challenge{DC5 -- Creation of training datasets for Bayesian 
networks}{To make the employed Bayesian network accurate, each topic 
must be provided with a similar number of training projects; otherwise, the 
obtained results can be strongly affected by the unbalanced distribution of the 
considered topics.}
\vspace{-.2cm}
\noindent
To mitigate such issues, we decided to train and evaluate the approach by 
considering 134 \GH featured topics. In this respect, we analyzed 13,400 README 
files by considering 100 repositories for each topic. To collect such 
artifacts, we needed to be aware of the constraints imposed by the \GH API, 
which limit the total number of requests per hour to 5,000 for authenticated 
users and 60 for unauthorized requests.

\smallskip
\noindent
\challenge{DC6 -- Usage of reusable heterogeneous components to 
develop new recommendation systems}{The orchestration of 
heterogeneous components was another challenge related to the development of 
\bnetwork.} 
\vspace{-.2cm}
\noindent
Though the employed Python libraries are flexible, they involve managing different technical aspects, \ie handling access to Web 
resources, textual engineering, and language prediction. Moreover, each component has a well-defined set of input elements that 
dramatically impact on the outcomes. For instance, the \RM encoding phase cannot occur without the data provided by the crawler 
component, which gets data from \GH. In the same way, the topic prediction component strongly relies on 
the feature extraction performed by the TF-IDF weighting scheme.
Thus, we succeeded in developing \bnetwork~by putting significant efforts in composing all the mentioned components coherently.

\smallskip
To summarize, concerning Fig. \ref{fig:RecsysFeatures}, NLP techniques have been applied to support the data preprocessing 
phase of \bnetwork. A model-based approach consisting of a Bayesian network underpins the overall technique to produce recommendations. 
The user 
context consists of an input 
README file, which is mined employing a keyword extraction phase. The 
produced recommendations are shown to the user directly in the employed 
Eclipse-based IDE.

\subsection{Lessons learned\label{sec:designLessons}}

%from the development of the \projectName \rs

Developing the \CM \rs has been a long journey, which allows us to 
garner many useful experiences. These experiences are valuable resources, which 
we can rely on in the future whenever we are supposed to run similar projects.

\smallskip
\challenge{DLL1 -- Selecting the right representation can be of 
paramount importance}{A suitable encoding helps capture the intrinsic features of the OSS ecosystem to facilitate the computation of similarities among software projects, moreover it paves the way for various recommendations.}
\vspace{-.2cm}
\noindent
With respect to the features shown in Fig.~\ref{fig:RecsysFeatures}, 
the used graph representation facilitates different recommendations, \ie \CS, 
\CR, 
and \FC making use of a graph-based	representation for supporting the 
\textit{Data Preprocessing} activity. We selected such a representation since some of the co-authors have gained similar experiences in the past and consequently, 
we followed the intuition to try with the adoption of graphs, and graph-similarity algorithms also in the mining OSS repositories. We started with \CS, and subsequently we found that the graph-based representation is also suitable to develop \CR and \FC.

\challenge{DLL2 -- Do not pretend to immediately find the optimal 
solution (move on iteratively)}{Conceiving the right techniques and 
configurations to satisfy user needs can be a long and iterative process.}
\vspace{-.2cm}
\noindent
For conceiving all the \rs we developed in \projectName, we followed an iterative process aiming to find the right underpinning 
algorithms and configurations to address the considered problems with the expected accuracy. It can be a very strenuous and Carthusian 
process that might require some stepping back if the used technique gives evidence of inadequacy for the particular problem at hand, 
fine-tune the used methods, and collect more data both for training and testing. For instance, in the case of \CS we had to make four 
main iterations to identify the features of open source projects relevant for solving the problem of computing similarities among 
software projects. During the initial iterations, we encoded more than the necessary metadata. For instance, we empirically noticed 
that encoding information about developers contributing to projects might reduce the accuracy of the proposed project similarity 
technique.

\smallskip
\challenge{DLL3 -- Start first with the techniques you already know and move on from there}{Starting with low-hang fruits allowed us to gain experience of the domain of interest and to quickly produce even sub-optimal results that can still be encouraging while finding better solutions.}
\vspace{-.2cm}
\noindent
During the execution of \projectName, we had a tight schedule, and once we agreed with the requirements by our use case partners, we started first with approaching 
problems that were similar to those we had already in the past. In other words, we first got low-hang fruits and then moved on from them. In this respect, we began early with \CS since we noticed some similarities with the problem that one of the co-authors previously addressed \cite{10.1145/2740908.2742141}. In this way, we managed to gain additional expertise and knowledge in the domain of \rs, while still satisfying essential requirements elicited from the partners. Afterwards, we succeeded in addressing more complicated issues, \ie recommending third-party libraries with \CR~\cite{Nguyen:2019:JSS:CrossRec} and API function calls and code snippets with \FC~\cite{Nguyen:2019:FRS:3339505.3339636,9359479}.

%% file: src/evaluation.tex
Once the \rs had been realized, it was necessary to compare them with 
existing state-of-the-art techniques. Evaluating a recommendation system is a 
challenging task since it involves identifying different factors. In 
particular, there is no golden rule for evaluating all possible 
\rs due to their intrinsic features as well as heterogeneity. To 
evaluate a new system, various questions need to be 
answered, as they are listed as follows:

%\color{blue}
\begin{itemize}
	\item [--] \emph{Which evaluation methodology is suitable?} Assessing RSSE can be done in different ways. Conducting a user study 
	has been accepted as the \emph{de facto} method to analyze the outcome of a recommendation process by several 
	studies~\cite{McMillan:2012:DSS:2337223.2337267,Moreno:2015:IUT:2818754.2818860,DBLP:journals/ese/PonzanelliBPOL16,10.1109/SANER.2017.7884605,10.1007/978-3-642-03013-0_15}.
	 However, user studies are cumbersome, and they may take a long time to finish. Furthermore, the quality of a user study's outcome 
	depends very much on the participants' expertise and willingness to participate. In this sense, setting up an automated evaluation, 
	in which the manual intervention is not required (or preferably limited), is greatly helpful. %Among others, we managed to avoid 
	%performing user studies by employing the \emph{k-fold cross validation} technique.
	\item [--] \emph{Which metric(s) can be used?} Choosing suitable metrics accounts for an important part of the whole evaluation process. 
	%for evaluating a recommender system has to be done by considering several aspects~\cite{10.1145/1864708.1864761, olmo_evaluation_2008}, including the datasets that are available, the existing baselines (if any), and the employed evaluation methodology. 
	While accuracy metrics, such as \emph{success rate}, \emph{precision} and \emph{recall} have been widely used to measure the 
	prediction performance, 
	we suppose that additional metrics should be incorporated into the evaluation~\cite{10.1145/1864708.1864761,Nguyen:2019:JSS:CrossRec}, aiming to study RSSE better. %it is not enough for studying all the performance traits of the recommendation outcomes, as it is the case with conventional recommender systems.	
	\item [--] \emph{How to prepare/identify datasets for the evaluation?} One needs to take into account different parameters when it 
	comes to choosing a dataset for evaluation. Moreover, the data used to evaluate a system depends very much on the underpinning 
	algorithms. %, among others. 
	%In fact, curating a dataset for evaluating a recommender system requires significant effort. 
	In this sense, advanced techniques and methods for curating suitable data are highly desirable. %of paramount importance; %For instance, it is necessary to perform various preprocessing steps to clean and refine the resulting data, such as filtering, manipulation and aggregation;
	\item [--] \emph{What could be a representative baseline for comparison?} To show the features of a new conceived tool and give 
	evidence of its novelty and advantages, it is necessary to compare it with existing approaches with similar characteristics. Since 
	the solution space is vast, comparing and evaluating candidate approaches can be a daunting task. %In fact, choosing the correct 
	%baseline is a challenging activity, %so as to ensure a fair comparison, the baselines to be considered must be endowed with 
	%%reusable \textit{tools} and \textit{datasets}. %When realizing a new type of recommendation system, it is crucial to have a clear 
	%%%knowledge of the possible techniques and patterns that might be employed to develop new ones. 
	
	%	\item [--] \emph{How do the current settings impact on the prediction performance?} Configuration a recommendation system depends on different aspects, \eg the size of the dataset, the kind of recommendation  the time based requirement, the employed recommendation algorithm must be adapted accordingly by changing its internal configurations. This can rely on different parameters that change according to the nature of the algorithm and the dataset used for .
\end{itemize}
%\color{black}
%	\begin{itemize}
%		\item Which evaluation methodology is 
%		suitable?
%		\item Which metric(s) can be used?
%		\item Which dataset is 
%		eligible/available for evaluation?
%		\item Which baseline(s) can be 
%		compared with?
%%		\item How to perform the evaluation?
%%		\item Which metrics need to be 
%%employed?
%%		\item Which baselines can be used for 
%%showing the novelties of 
%%the 
%%		proposed approaches?
%%		\item Which datasets can be used?
%	\end{itemize}

Answering such questions gives place to different challenges as 
described in the following subsection.

\vspace{-.5cm}

\subsection{Challenges}

\challenge{EC1 -- Identification of the suitable evaluation 
methodolgy}{Deciding the evaluation methodolgy to be applied has to 
take into account several aspects including the time and efforts that have been 
allocated for such a phase in the project the work is contextualized.}
\vspace{-.1cm}
\noindent
%Conducting a user study has been accepted as  the \emph{de facto} way to 
%analyze the outcome of a recommendation process by several studies 
%\cite{McMillan:2012:DSS:2337223.2337267,Moreno:2015:IUT:2818754.2818860,DBLP:journals/ese/PonzanelliBPOL16,10.1109/SANER.2017.7884605,10.1007/978-3-642-03013-0_15}.
%
User studies can be done as \emph{field studies} or as \emph{controlled 
	experiments}. By the former, the participants with different programming experience levels have to complete a list of tasks using 
	the proposed recommendation system without any intervention. The latter is conducted in a 
monitored environment, and the assigned tasks are carefully tailored for specific purposes. Although these strategies produce 
remarkable results in various work, there are some issues to be tackled. Among others, the selection of the participants has a crucial 
role to play.

It is worth noting that the 
selection of ground-truth data 
from an 
active project impacts the evaluation, and 
it might jeopardize the 
integrity 
of the evaluation process. Different aspects, 
\ie the scope of the 
recommendation, the recommendation input, the 
size of the ground 
truth, and the 
characteristics of the selected objects, 
should be carefully 
considered to 
mimic a real usage scenario when it comes to 
an automatized 
evaluation. For 
instance,  randomly choosing the ground truth 
size and objects does 
not 
guarantee that the evaluation mimics a real 
usage scenario. The ground 
truth 
extraction strategies that have been employed 
for evaluating the 
CROSSMINER \rs 
are explained below.

\smallskip
\noindent
\textit{\CR:} Given a set of libraries that an active project uses, \CR returns a set of additional libraries that similar projects to 
the active project have also included. For this reason, in the \CR evaluation process, given an active project, a half of its libraries 
are used as the ground-truth, and the remaining are used as the query. In this case, we split randomly into such sets the libraries 
that an active project includes.

\smallskip
\noindent
\textit{\FC:} Given a list of method declaration and method invocations pairs, and an active method context, \FC predicts the next 
method invocations that can be added to the active declaration. To simulate a developer's behaviour at different stages of a 
development project, 
we performed various evaluation experiments by varying the size of the recommendation query and the size of the ground-truth data. In 
particular, four different configurations have been considered in the evaluation to mimic the following scenarios:
	\begin{itemize}
		\item the developer is at an early 
		stage of the development 
		process, 
		and the active method is almost empty;
		\item the developer is at an early 
		stage of the development 
		process, 
		and the active method implementation 
		is well defined;
		\item the developer is near to the 
		end of the development 
		process, and 
		the active method is almost empty;
		\item the project is in an advanced 
		development phase, and the 
		active 
		method implementation is well defined.
	\end{itemize}
	The ground truth data is extracted 
	accordingly to the scenario 
	that the 
	evaluation mimics.
	
\smallskip
\noindent
\textit{MNBN:} Given an active 
	project, the MNBN recommendation 
	system 
	uses the content of \textit{README} 
	file(s) to recommend relevant 
	\GH 
	topics. Since the recommendation input 
	does not coincide with the 
	object of 
	the recommendation, we used the whole 
	list of topics that an 
	active project 
	uses as ground-truth.

%As previously mentioned, the  methodology  above can be applicable for 
%evaluating various recommendations. It has been suitable for the 
%following types of recommendations: third-party libraries 
%\cite{Nguyen:2019:JSS:CrossRec}, API function 
%calls~\cite{Moreno:2015:IUT:2818754.2818860,Nguyen:2019:FRS:3339505.3339636},
%and StackOverflow posts \cite{DBLP:journals/ese/PonzanelliBPOL16}. 
%Additional evaluations may also be possible depending on 
%the presence of input data. 

\smallskip
It is our firm belief that user 
studies are inevitable in 
many 
contexts. For the evaluation of \CS, a user 
study is a must, since 
there are no 
other ways to evaluate the similarity between 
two OSS repositories, 
rather than 
the manual scoring done by humans. We may 
avoid user studies in some 
specific 
cases. For instance, when evaluating \CR, we 
realized that with the 
application 
of the ten-fold cross-validation technique, 
we can rely on the 
available data 
to perform the evaluation, without resorting 
to a user study. 
%However, 
%we still 
%believe a user study makes sense for 
%evaluating \CR, since it allows 
%us to 
%better understand the usability of our 
%proposed approach. 
For \FC, 
while we can 
use data to evaluate its performance, we 
assume that its usability and 
usefulness can be properly studied only with 
a user study, where 
developers are 
asked to give their opinion on a specific API 
call recommended by the 
system.

%\item Which methodology?(\ie cross-fold 
%validation, user evaluation)
%\item \JDR{Uses evaluation challenges:

%\nb{Davide}{Check the coverage of the 
%following remaining items}
%	\begin{itemize}
%		\item Selection of the quality focus 
%by answering the 
%following 
%question: what does the user evaluation aim 
%to evaluate and how?
%		\item Selection of the evaluation 
%methodology, \ie an 
%empirical 
%evaluation assessment or a controlled 
%experiment where the 
%participants 
%perform 
%a set of tasks first, then they evaluate the 
%recommender system's 
%performance 
%by answering a questionnaire;
%		\item Designing the questionnaire and 
%decision of the 
%corresponding 
%analysis. For instance, open questions imply 
%a qualitative analysis, 
%whereas 
%closed or valuable questions could induce a 
%quantitative analysis;
%		\item Selection of context in term of 
%evaluators: the number 
%of 
%participants and their expertise, and 
%objects where the user study is 
%applied.
%\end{itemize}

\smallskip
\challenge{EC2 -- Selection of the right evaluation 
metrics}{Deciding which evaluation metrics %have to be employed 
for evaluating new \rs has to be done by considering several aspects, including the 
datasets that are available, 
the existing baselines (if any), and the employed evaluation methodology.}

%Given a query, the outcome of the 
%recommendation process is a ranked 
%list of 
%items that are considered to be relevant for 
%the query. For instance, 
%a system 
%that recommends third-party libraries for a 
%given project returns a 
%list in 
%descending order of similarity scores 
%corresponding to libraries 
%\cite{6671293}. To validate the performance 
%of a recommender system, 
%we need 
%a \emph{training} and a \emph{testing} 
%dataset 
%\cite{Cremonesi:2008:EMC:1468165.1468327}. 
%The former is used to 
%build the 
%model whereas the latter is used to validate 
%the outcome. Considering 
%a 
%project 
%that needs library recommendation, the graph 
%model is used to compute 
%similarities and then to find most $k$ 
%similar projects. , API 
%function 
%calls~\cite{Moreno:2015:IUT:2818754.2818860,Nguyen:2019:FRS:3339505.3339636}

\vspace{-.2cm}
\noindent
The recommendation outcome is normally a ranked list of items, \eg 
third-party libraries~\cite{Nguyen:2019:JSS:CrossRec,6671293}, API 
calls~\cite{Moreno:2015:IUT:2818754.2818860,Nguyen:2019:FRS:3339505.3339636},
or \GH topics~\cite{10.1145/3383219.3383227}. 
Normally, a developer pays attention only to the \emph{top-N} 
items. Thus, by comparing the items in the ranked list 
with those stored as ground-truth 
data, we can examine how well the recommendation system performs. There are various metrics to analyze the performance of a 
recommendation system. To our knowledge, several studies in RSSE focus only on 
accuracy~\cite{Bruch:2008:ERS:1454247.1454254,Fowkes:2016:PPA:2950290.2950319,McMillan:2012:DSS:2337223.2337267,6671293}.
However, in the scope of \CM, we realized that while accuracy is a good metric for evaluating an RS, it is not enough for studying all 
the performance traits of the outcomes, as it is the case with conventional recommendation systems 
\cite{10.1145/1864708.1864761}. As a result, other metrics should also be incorporated to analyze various quality aspects as they are 
presented as follows. First, the following notations are defined:

%To measure \emph{accuracy}, 
%\emph{precision}, \emph{recall} are 
%normally 
%utilized \cite{DBLP:conf/iir/NguyenRR18}. 
%\revised{Several studies pay attention only 
%to accuracy. However, in 
%the scope 
%of \CM, we realized that.}

%Among others, \emph{diversity}, 
%\emph{novelty} are popular metrics 
%for 
%studying 
%recommender systems. \emph{Diversity} means 
%the ability of the system 
%to 
%suggest to projects as many items as 
%possible, as well as to disperse 
%the 
%concentration among all the items. Whereas 
%\emph{Novelty} measures if 
%the 
%system is able to expose items to the 
%audience. 

\begin{itemize}
	\item \emph{N} is the cut-off value for 
	the list of recommended 
	items; %and 
	%\emph{k} is the number of neighbor 
	%projects considered for the 
	%recommendation process;
	\item for a testing project \emph{p}, the 
	ground-truth dataset is 
	named as 
	\emph{GT(p)};	
	\item $REC(p)$ is the \emph{top-N} items, 
	it is a ranked list in 
	descending 
	order of real scores, with $REC_r(p)$ 
	being the item in the 
	position 
	$r$;		
	\item if a recommended item $i \in 
	REC(p)$ for a testing project 
	$p$ is 
	found in the ground truth of $p$, \ie 
	\emph{GT(p)}, hereafter we 
	call this 
	as a \textit{match} or \textit{hit}.	
\end{itemize}

The metrics that have been employed for 
evaluating the \CM \rs are 
explained 
below.

%Using this notation, the metrics utilized to 
%measure the 
%recommendation 
%outcomes are explained below. Among others, 
%we consider \emph{success 
%rate} 
%\cite{6671293}, \emph{accuracy} 
%\cite{McMillan:2012:DSS:2337223.2337267}, 
%\emph{sales diversity}, and \emph{novelty} 
%the most suitable metrics 
%for 
%evaluating a recommender system in mining 
%OSS repositories 
%\cite{robillard_recommendation_2014}.% and 
%%they are recalled as 
%%follows.%, in 
%%%the context of. 

\smallskip
\noindent
\textit{\underline{Success rate.}} 
\label{sec:SuccessRate} Given a set 
of 
\emph{P} 
testing projects, this metric measures the 
rate at which a system can 
return at 
least a match among \emph{top-N} recommended 
items for every project 
$p \in P$ 
\cite{6671293}. It is formally defined as 
follows:

\begin{equation} \label{eqn:RecallRate}
success\ rate@N=\frac{ count_{p \in P}( \left 
|  GT(p) \bigcap 
(\cup_{r=1}^{N} 
REC_{r}(p)) \right | > 0 ) }{\left | P \right 
|} 
\end{equation}

\noindent
where the function \emph{count()} counts the 
number of times that the 
boolean 
expression specified in its parameter is 
\emph{true}. %Success rate 
%has been 
%mainly used to evaluate systems for 
%recommending similar software 
%systems/projects 
%\cite{10.1109/APSEC.2004.69,McMillan:2012:DSS:2337223.2337267}

\medskip
\noindent
\textit{\underline{Accuracy.}} Given a list 
of \emph{top-N} items, 
\emph{precision@N}, \emph{recall@N}, and 
\textit{normalized discounted cumulative gain 
(nDCG)} are utilized to 
measure 
the 
\emph{accuracy} of the recommendation results.

\textit{Precision@N} is the ratio of the 
\emph{top-N} recommended 
items 
belonging to the ground-truth dataset:

\begin{equation} \label{eqn:Precision}
precision@N(p) = \frac{\sum_{r=1}^{N}\left |  
GT(p) \bigcap REC_{r}(p) 
\right 
|}{N}
\end{equation}

\textit{Recall@N} is the ratio of the 
ground-truth items appearing in 
the 
\emph{N} items 
\cite{Davis:2006:RPR:1143844.1143874,DiNoia:2012:LOD:2362499.2362501,Nguyen:2015:CRV:2942298.2942305}:
 %,ISINKAYE2015261, 

\begin{equation} \label{eqn:Recall}
recall@N(p) = \frac{\sum_{r=1}^{N}\left |  
GT(p) \bigcap REC_{r}(p) 
\right 
|}{\left | GT(p) \right |}	
\end{equation}

\textit{nDCG:} Precision and recall reflect 
well the accuracy, however 
they 
neglect ranking sensitivity 
\cite{BellogiN:2013:CSH:2397740.2398191}. 
nDCG is an effective way to 
measure if a system can present highly 
relevant items on the top of 
the list:%. 
%.measure the sensitivity of a ranking scheme 
%For binary ratings, the 
%following 
%formula is used to computed the metric. %It 
%%incorporates.
%\cite{DINOIA2017234}

\begin{equation} \label{eqn:nDCG}
nDCG@N(p) = \frac{1}{iDCG} 
\cdot\sum_{i=1}^{N} 
\frac{2^{rel(p,i)}}{log_{2}(i+1)}
\end{equation}

where iDCG is used to normalize the metric to 
$1$ when an ideal 
ranking is 
reached. %The ranking

%As accuracy is not sufficient to evaluate 
%recommender systems 
%\cite{McNee:2006:AEA:1125451.1125659}, we 
%also consider Sales 
%Diversity and 
%Novelty as follows.
%, thus increasing the chance that products 
%will get sold by being 
%introduced

\medskip
\noindent
\textit{\underline{TopRank.}} It measures the 
percentage of the first 
top 
elements 
in the ground-truth data:
%\vspace{-.1cm}
\begin{equation} \label{eqn:topRank}
Top\ rank  = \frac{ TpRank(r) }{\left | R 
\right |}    \times 100\%
\end{equation}
where \emph{TpRank(r)} returns 1 if the first 
predicted element 
belongs to 
\textit{UsrTp(r)}, 0 otherwise.

\medskip
\noindent
\textit{\underline{Sales Diversity.}} In 
merchandising systems, 
\emph{sales 
diversity} 
is the ability to distribute the products 
across several customers 
\cite{Nguyen:2015:CRV:2942298.2942305,Vargas_sales_diversity_14}.
 In 
the 
context of mining software repositories, 
sales diversity means the 
ability of 
the system to suggest to projects as much 
items, \eg libraries, code 
snippets, 
as possible, as well as to spread the 
concentration among all the 
items, 
instead of presenting a specific set of them 
\cite{robillard_recommendation_2014}. %To 
%gauge \emph{sales 
%diversity}, we 
%utilize \emph{catalog coverage} and 
%\emph{entropy}.% 
%%\cite{Nguyen:2015:ESP:2740908.2742141}.

\noindent
\textit{Catalog coverage} measures the 
percentage of items recommended 
to 
projects:%\cite{Nguyen:2015:ESP:2740908.2742141}:
% %indicates how well 
%%a 
%%recommender system is able to: %expose 
%%%libraries to projects: 

\begin{equation}\label{eqn:Coverage}
coverage@N = \frac{\left | \cup_{p\in P} 
\cup_{r=1}^{N} REC_{r}(p) 
\right | 
}{\left | I \right |} 
\end{equation}	

\noindent 
where $I$ is the set of all items available 
for recommendation and $P$ 
is 
the set of projects.

\noindent
\textit{Entropy} evaluates if the 
recommendations are 
concentrated 
on 
only a 
small set or spread across a wide range of 
items: 
%\cite{Ragone:2017:SLF:3019612.3019837}: 
%%%measures the distribution of 
%%the 
%recommendations across all the items, 
%showing whether the 
%recommendations are 
%concentrated on a few items or are better 
%distributed:

\begin{equation}\label{eqn:Entropy}
entropy = -\sum_{i \in I}\left ( 
\frac{\#rec(i)}{total} \right )ln 
\left ( 
\frac{\#rec(i)}{total} \right )
\end{equation}

\noindent
where $\#rec(i)=count_{p \in 
P}( \left |   (\cup_{r=1}^{N} REC_{r}(p)) \ni 
i  \right |  )$, ($i \in 
I$)  is 
the number of projects getting $i$ as a 
recommendation, %$I$ is the 
%set of all 
%items available for recommendation, 
$total$ denotes the total number of 
recommended items across all 
projects.

%in this sense, a lower entropy value 
%indicates a better distribution 
%and vice 
%versa.	

%\item Gini Index.

%have never been recommended
%Expected popularity complement (EPC). We 
%apply the population-based 
%item 
%novelty evaluation metric proposed in [31] 
%which is called. to 
%measure the 
%ability of a recommender system to recommend 
%items from the long tail 
%As Anderson [4] stated, some businesses or 
%economic models present a 
%Long tail 
%effect, in which a few of the most popular 
%items are extremely 
%popular, while 
%the rest – the long tail – is much less 
%known. Promoting the 
%recommendation of 
%items in this long tail may offer benefits 
%for both users and the 
%business 
%behind the recommender system 
%\cite{Vargas_sales_diversity_14}.

%The Long Tail theory suggests that, as the 
%Internet makes 
%distribution easier 
%and technology systems allow consumers to 
%become aware of more 
%obscure 
%products, demand will shift from the most 
%popular products at the 
%“head” of a 
%demand curve to the aggregate power of a 
%long “tail” made up of 
%demand for 
%many 
%different niche products
% Figure \ref{fig:NumOfPros} reflects the 
%long tail effect.  we can 
%see the 
%long tail effect

%can recommend unpopular libraries, i.e. 
%those that are much less 
%exposed to 
%projects

%,Shi:2014:CFB:2620784.2556270
%The long tail effect can be spotted in 
%Figure \ref{fig:NumOfLibs} as 
%a few of 
%libraries are very popular by being included 
%in several projects, 
%whereas most 
%of the libraries appear in much less 
%projects. 

\medskip
\noindent
\textit{\underline{Novelty.}} The metric 
%
%In business, t
%The \emph{long tail effect} is the fact that 
%a few of the most 
%popular 
%products are extremely popular, while the 
%rest, so-called the long 
%tail, is 
%obscure to customers 
%\cite{Anderson:2006:LTW:1197299}. It is 
%expected 
%that 
%products in the long tail can also be 
%introduced, since this is 
%beneficial to 
%both customers and business owners 
%\cite{Vargas_sales_diversity_14}. 
%By library recommendation, 
gauges if a system is able to expose items to 
projects. 
\emph{Expected popularity complement} (EPC) 
is utilized to measure 
\emph{novelty} and is defined as follows 
\cite{Vargas:2011:RRN:2043932.2043955,Vargas_sales_diversity_14}:
 
%This might 
%help developers find \emph{serendipitous} 
%libraries, e.g. those that 
%are seen 
%by chance but turn to be useful for their 
%current project. 
%Niemann:2013:NCF:2487575.2487656,

\begin{equation}\label{eqn:EPC}
EPC@N = \frac{\sum_{p\in P}\sum_{r=1}^{N} 
\frac{ rel(p,r)* \left [ 
1-pop(REC_{r}(p)) \right ]}{log_{2}(r+1)} 
}{\sum_{p\in 
P}\sum_{r=1}^{N} 
\frac{rel(p,r)}{log_{2}(r+1)}}
\end{equation}

\noindent
where $rel(p,r)=\left |  GT(p) \bigcap  
REC_{r}(p) \right |$ 
represents the 
relevance of the item at the $r$ position of 
the \emph{top-N} list to 
project 
$p$; $pop(REC_{r}(p))$ is the popularity of 
the item at the position 
$r$ in the 
\emph{top-N} recommended list. It is computed 
as the ratio between the 
number 
of projects that receive $REC_{r}(p)$ as 
recommendation over the 
number of 
projects that are recommended items among the 
most often recommended 
ones. 
Equation \ref{eqn:EPC} implies that the more 
unpopular items a system 
recommends, the higher the EPC value it 
obtains and vice versa. 
%TODO check if it should be included
%\subsection{Average Reciprocal Hit-Rank}
%\emph{Average Reciprocal Hit-Rank (ARHR)} is 
%a common index to 
%evaluate the 
%ranking of top-n recommended items. It is 
%particular suitable when 
%evaluating 

%It is expected that a recommender system can 
%recommend as much as 
%possible 
%items in the long tail.
%In this sense, a system that recommends items
% \cite{Niemann:2013:NCF:2487575.2487656}

\medskip
\noindent
\textit{\underline{Confidence.}} Given a pair 
of $<$\textit{query, 
retrieved 
item}$>$ confidence is the score the 
evaluator assigns to the 
similarity 
between the two items;

%the ranking calculated by the similarity 
%tools and 
%Because of the large number of ties, w

\medskip
\noindent
\textit{\underline{Ranking.}} In a ranked 
list, it is necessary to 
have a good 
correlation with the scores given by the 
human 
evaluation~\cite{Bruch:2008:ERS:1454247.1454254}.
 The Spearman's rank 
correlation coefficient $r_{s}$ 
\cite{spearman1904proof} is used to 
measure how 
well a similarity metric ranks the retrieved 
projects given a query. 
Considering two ranked variables 
$r_{1}=(\rho_{1},\rho_{2},..,\rho_{n})$ and 
$r_{2}=(\sigma_{1},\sigma_{2},..,\sigma_{n})$,
 $r_{s}$ is defined as: 
$r_{s}=1-\frac{6\sum_{i=1}^{n} 
(\rho_{i}-\sigma_{i})^{2}}{n(n^{2}-1)}$. We 
also 
employed \emph{Kendall's tau} 
coefficient~\cite{kendall1948rank}, 
which is used 
to measure the ordinal association between 
two considered quantities. 
Both 
$r_{s}$ and $\tau$ range from -1 (perfect 
negative correlation) to +1 
(perfect 
positive correlation); $r_{s}=0$ or $\tau=0$ 
implies that the two 
variables are 
not correlated.% %; $r_{s}$ ranges from 
%%$-1.00$ (perfect negative 
%%correlation) 
%%and $+1.00$ (perfect positive correlation)

\medskip
\noindent
\textit{\underline{Recommendation time.}} 
Being able to provide 
recommended 
items 
in a limited amount of time is important, 
especially for applications 
that 
require instant recommendations. This metric 
evaluates the duration of 
time, 
starting from when a user sends the query 
until the final 
recommendations are 
returned.
%As mentioned in \textbf{RQ$_2$}, we measure 
%the time needed by both 
%PAM and 
%FOCUS to perform a prediction on a given 
%infrastructure, which is a 
%laptop 
%with 
%Intel Core i5-7200U CPU @ 2.50GHz $\times$ 
%4, 8GB RAM, and Ubuntu 
%16.04.

\smallskip
Depending on the context, we have to choose a suitable set of 
metrics to evaluate a recommendation system. For example, with \CS we can only make use of \textit{Success rate}, \textit{Confidence}, 
\textit{Precision}, \textit{Ranking}, and \textit{Execution time} to evaluate the tool,  since we relied on a user study. Meanwhile, 
with \CR or \FC, since we can use the ten-fold cross validation technique (i.e., by exploiting the testing data which was already split 
into query and ground-truth data), we evaluated them using \textit{Accuracy}, \textit{Precision}, \textit{Recall}, \textit{Diversity}, 
and \textit{Novelty}.

%\nb{Davide}{The following are missing: long 
%tail, statistics tools}

\smallskip
\challenge{EC3 -- Identification of the datasets that are 
eligible and available for evaluation}{The 
datasets used for evaluation very much depend 
on the recommendation 
algorithms being used.}
\vspace{-.2cm}
\noindent
For each developed tool, we had to go through the 
following dimensions related to datasets:

%	\begin{itemize}	
%	\item Which dataset? It depends on the 
%algorithms being used;
	\begin{itemize}
%		\item \textit{Availability of the 
%datasets}, \eg metadata, 
%source code, 
%communication channel 
%		data, version control
%		systems information, bug tracking 
%system messages and errors, 
%continuous integration information, 
%developing workflow as pull request, \etc;
		 \item \textit{Which format?} 
		 Depending on the employed 
		 recommendation 
		 techniques (\eg 
		 collaborative filtering, CNNs, etc.) 
		 we had to identify the 
		 proper 
		 ways to encode the created 
		 datasets. For instance,  to enable 
		 the application of a 
		 graph-based 
		 similarity algorithm 
		 underpinning \CS, we had to encode 
		 the different features of 
		 OSS 
		 projects in a graph-based 
		 representation. The same datasets 
		 needed to be represented in 
		 a TF-IDF 
		 format to enable the 
		 application of \FC;
		
		\item \textit{Which preprocessing 
		process should be applied to 
		create 
		the dataset?} To minimize the size of 
		the input datasets and 
		thus to 
		make their manipulation efficient, we 
		had to perform different 
		data 
		filtering tasks. For instance, in the 
		case of \CS, to enable 
		the 
		application of the employed 
		graph-similarity algorithm, we 
		identified 
		the features that are relevant for 
		the task. For example, 
		information 
		about software developers, source 
		code, and \GH topics was 
		filtered out 
		from the available datasets even 
		though it was easy to encode 
		all of 
		them as elements in the input graphs. 
		Similar data filtering 
		phases 
		were also performed in \CR to enable 
		the recommendation of 
		third-party 
		libraries that might be added in the 
		project under 
		development.  
		Indeed, such data filtering phases 
		have to be performed 
		without 
		compromising the performance (in 
		terms of accuracy, precision, 
		recall, 
		etc.) of the approach under 
		evaluation;
		
		\item \textit{Which limitations 
		should we tackle  when 
		collecting the 
		dataset?} The primary 
		limitations we experienced when 
		evaluating the \projectName 
		\rs were 
		related to the \GH APIs 
		restrictions. Unfortunately, the 
		adoption of alternative 
		sources like 
		GHTorrent \cite{Gousi13} was not 
		enough due to the lack of 
		needed 
		artifacts such as source code. 
		Knowing such limitations in 
		advance, 
		when collecting projects from \GH, we 
		decided to save 
		as much data as possible for every 
		single project. The goal 
		was to 
		enable the reuse of the 
		collected data even for perspective 
		evaluations to be done for 
		future 
		\rs to be developed in the 
		context of \projectName.
		
%		\item \textit{Which Size?}
%		The size is a sneaky problem, in one 
%hand there is the 
%undisputed 
%advantage of relying on
%		large datasets, since more data means 
%more information, on the 
%other 
%hand we have to face
%		with a physical storage limitation 
%and computational issues. 
%The 
%dimension span from few Mb
%		to hundred of Gb. During the 
%evaluation of \CS we encountered 
%many 
%issues related to size, 
%		namely the dataset we used was too 
%big for MUDABlue, this led 
%to many 
%heap runtime errors.
%		We managed the problem by increasing 
%the heap dimension, but 
%this 
%solution could be always available.

	\end{itemize}	
%\end{itemize}

\smallskip
\challenge{EC4 -- Identification of 
the baseline(s) to 
compare with}{To showcase the 
features of a 
new conceived 
tool as well as to 
demonstrate its novelty, it is necessary to 
compare it with existing 
approaches 
with similar features. In fact, choosing the 
correct baseline is a 
challenging 
activity, as to ensure a fair comparison, the 
baselines to be 
considered must 
be endowed with reusable \textit{tools} and 
\textit{datasets}.}
\vspace{-.2cm}
\noindent
While in general, authors of the selected baselines published their 
tools and 
dataset online, it is the case that many of them are faulty, or not well 
maintained, or even worse: no longer available. In particular, while 
developing the \projectName \rs we always tried our best to identify the 
baselines to be used for the evaluations. Unfortunately, often they were 
not available, which indeed led to difficulties in the evaluation. For 
instance, for evaluating \CS, since the implementations of the 
baselines were no longer available for public use, we had to re-implement them  by strictly following the descriptions 
in the original papers 
\cite{10.1109/APSEC.2004.69,McMillan:2012:DSS:2337223.2337267,10.1109/SANER.2017.7884605}.
That was not possible for evaluating \bnetwork~due to the lack of details in the publicly available documents describing the 
corresponding baseline. In general, whenever a baseline is selected, and it is not available online, we contacted its authors for the 
original implementation. It was rare that we got a response from the authors with the tool and/or 
data. Thus, for the particular cases 
of the developed \rs, 
%we experienced some situations where the 
%baseline existed and it was adequately 
%described on papers, but the 
%related 
%software tools were not available. Given 
%that a tool cannot be 
%provided, there 
%are two possible ways to overcome the 
%difficulty. 
either we re-implemented the tool, as it is the case with \CS, 
%we had to fairly re-implement all the 
%baselines by following the details given in 
%the related papers. 
or we compared by performing experiments on the datasets 
that have been used in the original papers, as we did for \CR. 

\begin{table}[t!]
	\centering
	\footnotesize
	\scriptsize
	\vspace{-.2cm}
	\caption{\CM \rs: evaluation facts.}
	\vspace{-.2cm}
	\label{tab:EvaluationFacts}
	\begin{tabular}{|p{1.5cm} | p{1.5cm}| 
	p{1.6cm} | p{1.6cm} | 
	p{1.8cm} | 
	p{1.2cm} |}  \hline				
		\rowcolor{verylightgray}
		\multicolumn{2}{|c|}{ } & 
		\CS~\cite{8498236} & 
		\CR~\cite{Nguyen:2019:JSS:CrossRec} & 
		\FC~\cite{Nguyen:2019:FRS:3339505.3339636,9359479}
		 & 
		\bnetwork~\cite{10.1145/3383219.3383227}
		 \\ \hline \hline	
		\multirow{2}[1]{*}{Methodology} & 
		Cross-Val. &  & \cmark & 
		\cmark & 
		\cmark \\ \cline{2-6}		
		& User study & \cmark &  &  &  \\ 
		\hline \hline
		\multirow{11}[5]{*}{Metric} & Success 
		rate & \cmark &  &  &  
		\\ 
		\cline{2-6}		
		& Precision & \cmark & \cmark & 
		\cmark & \cmark \\ \cline{2-6}
		& Recall &  & \cmark & \cmark & 
		\cmark \\ \cline{2-6}
		& nDCG &  & \cmark &  &  \\ 
		\cline{2-6}
		& TopRank &  &  &  & \cmark \\ 
		\cline{2-6}				
		& Coverage &  & \cmark &  &  \\ 
		\cline{2-6}		
		& Entropy &  & \cmark &  &  \\ 
		\cline{2-6}		
		& Novelty &  & \cmark &  &  \\ 
		\cline{2-6}	
		& Confidence & \cmark &  &  &  \\ 
		\cline{2-6}		
		& Ranking & \cmark &  &  &  \\ 
		\cline{2-6}
		& Time & \cmark &  & \cmark &  \\ 
		\hline \hline	
		\multirow{2}[2]{*}{Dataset} & Source 
		& \GH & \GH & \GH, Maven 
		central 	
		repository& \GH \\ \cline{2-6}
		& Size & 580 projects  & 1,200 
		projects & 3,600 projects & 
		13,400 
		projects \\ \hline	\hline
		\multicolumn{2}{|l|}{Artifact} & 
		Metadata  & Metadata & Source 
		code & 
		\RM 
		files \\ \hline	\hline 
		\multicolumn{2}{|l|}{Baseline} & 
		MUDABlue~\cite{10.1109/APSEC.2004.69},
		 
		CLAN \cite{McMillan:2012:DSS:2337223.2337267},
		 
		RepoPal \cite{10.1109/SANER.2017.7884605}
		 & 
		LibRec~\cite{6671293}, 
		LibFinder~\cite{Ouni:2017:SSL:3032135.3032325},
		 
		LibCUP~\cite{SAIED2018164} & 
		UP-Miner~\cite{Wang2013Mining}, PAM~\cite{Fowkes:2016:PPA:2950290.2950319}
		 
		& None \\ \hline%\cline{2-6}
		%		& abc & abc  & abc & abc & 
		%abc \\ \hline			
	\end{tabular}
	\vspace{-.4cm}
\end{table}
% \footnotemark[2] The use of Rascal is 
%optional.

\medskip 
Table~\ref{tab:EvaluationFacts} summarizes 
the main factors related to the evaluation of our proposed recommendation systems. Depending on the intrinsic characteristics of each 
tool, different metrics and methodologies were employed to evaluate them. For example, to study 
\CS~\cite{8498236,Nguyen:2019:JSS:CrossSim}, 
a user study with several developers' involvement was the only option since there is no automated method to evaluate the similarity 
between two OSS projects. Meanwhile, with 
\CR~\cite{Nguyen:2019:JSS:CrossRec}, 
\FC~\cite{Nguyen:2019:FRS:3339505.3339636}, 
and \bnetwork~\cite{10.1145/3383219.3383227}, we 
relied only on data to investigate their performance. Moreover, depending on the availability of baselines and quality requirements, we 
used different evaluation metrics, such as Accuracy (Precision, Recall, TopRank) or Sales Diversity (Coverage, Entropy).  Choosing 
suitable data plays an important role in the evaluations, and it depends on various factors, such as systems' characteristics, 
baselines, 
evaluation purposes, or even constraints imposed by OSS platforms, \eg 
\GH and the Maven central repository. The selection of baselines was also a significant issue, considering their complexity and 
relevance with our tools. For evaluating \CR, we were able to consider three different tools for comparison, \ie 
LibRec~\cite{6671293}, LibFinder~\cite{Ouni:2017:SSL:3032135.3032325},
and LibCUP~\cite{SAIED2018164}. While with \FC, 
only PAM~\cite{Fowkes:2016:PPA:2950290.2950319} 
was selected to benchmark since other relevant tools such as MAPO~\cite{10.1007/978-3-642-03013-0_15} 
and UP-Miner~\cite{Wang2013Mining} were no longer available. In summary, we believe that there are many factors when it comes to design 
and evaluate a recommendation system, and we should carefully investigate the most probable scenarios to select the optimal one.

\subsection{Lessons learned} % by evaluating the \projectName \rs

\smallskip
\textbf{ELL1 -- User studies are cumbersome, and they can 
		take a long time to be conducted and completed:} 
		The quality of a user study's outcome depends very much on the participants' expertise and willingness to participate. People 
		are often not very keen on the experiments, since there is no incentive/reward for performing the required tasks. Moreover, 
		there is a trade-off between domain-expert developers, who may not need a recommendation system to develop, and students who 
		have never used this type of system. As a result, we evaluated \CS by involving 15 developers of different background of 
		knowledge. Aiming at a reliable evaluation, for each query we mixed and shuffled the top-$5$ results generated from the 
		computation by each similarity metric in a single Google form and presented them to the evaluators who then inspected and given 
		a score to every pair. Thus, we managed to mimic a \emph{taste test} where users are asked to judge a product, \eg~food or 
		drink, without 
		having a priori knowledge about what is being evaluated \cite{Ghose2001,doi:10.1108/13522750810879048}. In this way, we removed 
		any bias or prejudice against a specific similarity metric. The participants were asked to label the similarity for each pair 
		of query and retrieved project regarding their application domains and functionalities. Furthermore, we also allowed for 
		cross-checking, \ie the results of one developer were validated by the others. To perform such evaluation for \CS and compare 
		it with the baselines, it has been crucial to design the experimental settings properly and clearly define the manual 
		evaluation tasks by adhering to the taste-test methodology.

\smallskip
\noindent
\textbf{ELL2 -- In some certain contexts, the k-fold cross-validation technique is a good alternative to user studies:} As previously 
mentioned, through \CM, we realized that user studies are cumbersome, and they can take a long time to conduct and complete. 
However, we experienced that the assessment can also be automatized by means of case studies or data itself. By the former, use cases 
are pre-selected for the recommendation. By the latter, we set up an automated evaluation, in which the manual intervention is not 
required, or preferably limited. 
Depending on the availability of data, we managed to avoid performing user studies by employing the \emph{k-fold cross validation} 
technique \cite{WONG20152839}, which has been popularly chosen as the evaluation method for a model in Machine Learning. By this 
method, a dataset is divided into $k$ equal parts (\emph{folds}). For each 
validation round, one fold is used as a testing and the remaining \emph{k-1} 
folds are used as training data. Such an evaluation attempts to mimic a real scenario: \emph{the system should produce recommendations 
for a project based on the data available from a set of existing projects}. The artifact being considered as the recommendation target 
is called \emph{object}. For instance, 
regarding third-party libraries recommendation  
\cite{Nguyen:2019:JSS:CrossRec,6671293}, 
objects are libraries that a system provides as its outcome. It is essential to study if the recommendation system is useful by 
providing the active project with relevant libraries, exploiting the training data. To this end, we keep a certain amount of objects 
for each active project and use them as input for the recommendation engine, which can be understood as the query. The rest is taken 
out and used as \emph{ground-truth data}. The 
ground-truth data is compared with the recommendation outcomes to validate the system's performance. It is expected that the 
recommendation system can retrieve objects that match up the ones stored as ground-truth data.

\smallskip
\noindent
\textbf{ELL3 -- The quality of data 
depends on the particular 
application domain of interest:} %By working on the \projectName \rs, 
Through \projectName, we further confirmed the importance of having the availability of big data and high-quality data for training and evaluation activities. The 
definition of data quality 
cannot be 
given in general, and it very much depends on 
the particular 
application of 
interest. According to our experience, 
creating a dataset, which can be rightly used 
for both training and 
evaluating 
the developed \rs, can require significant 
effort, which can be 
comparable to 
that needed to realize the conceived 
approach. For instance, to 
implement 
\bnetwork, we devoted a huge effort to create 
the dataset that was 
supposed to 
be 
balanced with respect to the considered \GH 
featured topics. Moreover, 
it can 
be challenging to collect big datasets, 
especially when there are 
several 
constraints to be satisfied. For instance, in 
the case of the \FC 
evaluation, 
one of the considered datasets was initially 
consisting of 5,147 Java 
projects 
retrieved from the Software Heritage archive 
\cite{SH}. To comply with
the requirements of the baseline, we first 
restricted the dataset to the list of 
projects that use at least one 
of the 
considered third-party libraries. Then, to 
comply with the 
requirements of \FC, 
we restricted the dataset to those projects 
containing at least one 
\code{pom.xml} file. Because of such 
constraints, we ended up with a 
dataset 
consisting of 610 Java projects. Thus, we had 
to create a dataset ten 
times 
bigger than the used one for the evaluation.

\smallskip
\noindent
\textbf{ELL4 -- Candidate baselines might not be reusable:} 
When conceiving new \rs there can be no baselines to compare with. There are at 
least two motivations: \emph{(i)} the proposed approach is the first attempt dealing 
with the considered problem; \emph{(ii)} the tools and datasets of existing baselines 
are no longer available or reusable. In such cases, according to the facts 
shown in Table \ref{tab:EvaluationFacts}, the k-fold cross-evaluation has been 
a valuable technique that allowed us to evaluate most of the proposed \rs even 
when the baselines were not available. Concerning \CS, we decided to perform a 
user study, to mitigate any bias related to the fact that we re-implemented all 
the baselines.

\smallskip
\noindent
\textbf{ELL5: Novelty and diversity are good indicators that 
are worth considering:} Many existing approaches just choose to recommend \textit{popular} items, \eg 
USE~\cite{Moreno:2015:IUT:2818754.2818860}, 	\PR~\cite{DBLP:journals/ese/PonzanelliBPOL16}, LibRec~\cite{6671293}. Through the 
evaluation of \CR, we demonstrated that further than popularity, \textit{novelty} and \textit{diversity} are good indicators for 
assessing if the recommendation outcomes are meaningful. 
Among others, the ability to recommend items in the long tail is essential: we can suggest things even when extremely unpopular since a 
small number of projects use each. However, they turn out to be useful as all of them match those stored as ground-truth. This implies 
that the novelty of a ranked list is important: a system should recommend libraries that are \emph{novel} \cite{Castells_noveltyand}, 
\ie those that have been rarely seen. In this sense, we see that \CR can produce good outcomes, not only in terms of success rate and 
accuracy but also \textit{sales diversity} and \textit{novelty}. Moreover, \textit{serendipity} has been widely exploited to evaluate 
\rs in other domains. Serendipity means that items are obtained by chance but turn out to be useful. However, it seems that the metric 
has been neglected in evaluating \rs in software engineering. Investigating the importance of serendipity in the context of source 
code/library recommendation can be an interesting topic. For example, a recommendation engine provides a developer with an artifact, 
\eg a third-party library or an API function call, which does not belong to the ground-truth data at all; however, it is indeed useful 
for the current project.

%According to evaluation facts shown in Table 
%\ref{tab:EvaluationFacts}, 
%\textit{precision} and \textit{recall} are 
%the metrics that we have 
%considered

%% file: src/RelatedWork.tex
In this section we provide a literature review %of existing studies %that have reported 
on the development and usage of \rs in software engineering. More importantly, we associate our work to various existing studies, aiming to highlight its main contributions.

%associate our work with to various existing studies
%It collects, structures and formalizes knowledge on recommendation systems in software engineering. 

In their book~\cite{robillard_recommendation_2014}, Robillard \etal focus on the techniques and applications of \rs in software engineering. The work presents a pragmatic approach to system design, implementation, and evaluation. Similarly, Proksch \etal~\cite{DBLP:conf/laser/ProkschBM14} present a comprehensive report on different phases that need to be considered when developing %must be taken to develop 
	an effective recommender system to support the development activities. %The work provides a ... by reporting various steps. %, implementation, as well as evaluation for. 
	Though these studies are highly related to our work, they provide a set of guidelines for developing and evaluating a generic recommendation system. In other words, such guidelines are not tailored to any specific recommendation systems. %If someone wants to look. 
	There is a lack of %experience there is a lack of 
	proper references for anyone who wants to customize their implementation for a specific context. For example, a developer is interested in understanding which techniques can be used for producing recommendations; %or how to capture the developer’s context; 
	or which evaluation metrics are suitable for the results obtained by conducting a user study. %the most feasible way to present recommendation outcomes, to name a few. %with a concrete way to deal with various issues, for instance . %experience report. 
	And this is where our work comes in: %\ie 
%	Meanwhile, in our work, 
	we %intended to %overcome such limitation 
	complement the existing studies by reporting on a specific use case, \ie \rs developed through the \projectName project to satisfy requirements imposed by various industrial partners. More importantly, we provide the community at large with detailed challenges and lessons. In this respect, our work is expected to be a practical benchmark, when it comes to design and implementation of a recommendation system for mining OSS repositories.  %we are able to

%By investigating various related work, we realized that while they tackled well the issue of designing and implementing a recommendation system for software engineering in general, there is a lack of proper references for anyone who wants to customize their implementation in a specific context. For example, a developer may be interested in understanding which data mining techniques are suitable for producing recommendations; or how to capture the developer’s context; or which is the most feasible way to present recommendation outcomes, to name a few.

%experiences about 
%shares some,
%	This book collects, structures and formalizes knowledge on recommendation systems in software engineering. It adopts a pragmatic approach with an explicit focus on system design, implementation, and evaluation. 

\smallskip
Pakdeetrakulwong \etal~\cite{pakdeetrakulwong_recommendation_2014} investigate the impact 
of \rs on the software development life cycle (SDLC). By analyzing several state-of-the-art studies, they identified three main components of a recommendation 
system, \ie a mechanism to collect data, a recommendation engine, and a user interface to deliver recommendations. A recommendation system should 
support a developer throughout the SDLC phases, ranging from design to testing. Among others, the most supported one by existing recommender 
tools is the implementation phase, in which software engineers turn components designed in a previous phase into code. Although implementation is a crucial phase in the software engineering (SE) domain, the other phases are required, \ie collecting requirements, design phase, and testing. To be more concrete, a recommendation system should be able to provide several types of items, including UML diagrams and artifacts. A promising filed is Semantic Web and Ontologies, which are used to describe software components in SE that allow information sharing among team members. Moreover, most of RSSEs use a pull approach to deliver recommendations: it means that recommended items are provided without any specific request made by the user. Reversely, the work~\cite{pakdeetrakulwong_recommendation_2014} suggests using a push approach, in which the user can trigger the delivery of the recommendations reactively.

%\smallskip
Maki \etal~\cite{maki_context} propose a feature model to represent the 
problem of capturing contexts in the RSSE domain. As a preliminary analysis, 
the authors discuss 23 papers and classify them in the following six categories:
\begin{itemize}
	\item \textit{Change task}: this type of recommendation system aims to support 
	the developer in managing the evolution of the current 
	programming task;
	\item \textit{API usage}: this type of RSSEs supports the usage of external 
	third-party libraries;
	\item \textit{Refactoring task}: \rs that support refactoring activities    
	fall in this category; 
	\item \textit{Solving exception, failure, and bug}: this kind of RSSEs 
	handles the exception and unexpected behaviours of the considered software 
	systems;
	\item \textit{Recommending software components and components' design}: it 
	recommends entire software components to be integrated into the software 
	projects under development;
	\item \textit{Exploring local codebases and visited source locations}: this type of system supports the information search over different data sources, \ie online datasets
\end{itemize}

In the same paper \cite{maki_context}, the authors show that the examined tools 
work in practice, but they fail to address the \textit{capturing context} phase accurately. Such a phase plays an essential role in the overall 
recommendation process, as it is performed at the early stages of the 
production. According to the authors \cite{maki_context} the context extraction 
phase should be triggered \emph{(i)} \textit{reactively} 
or \emph{(ii)} \textit{proactively}. The former is led by certain actions 
captured in the development environment, \eg page scrolling or considering idle 
times. 
Contrariwise, the latter is activated directly by the user. Then, the capturing 
phase is performed by setting the scope and the elements to be extracted. The 
scope dimension depends on the goal that the RSSE wants to achieve. Thus, 
recommendation systems can consider as snippets of code as the context as well 
as the entire project. Considering the possible extracted elements, they can be 
related to valuable elements of the code, \ie variables, methods or identifiers. 
Similarly to the previous phase, the treatment of the excepted data can be 
different according to the RSSE's aims. Standard techniques to perform this 
step involve parsing, weighting or filtering. The final step is the delivery of the recommendations, according to various output formats, \ie bag-of-words, 
AST, dendrograms, annotated graphs, and weighted vectors. The findings of the 
work suggest that most of the analyzed tools do not cover the context 
extraction activity properly. Thus, there are room of improvements in this 
filed that can bring substantial contribution in the RSSE domain. 

\smallskip
Happel \etal~\cite{happel_potentials_2008} discuss relevant issues that a 
recommendation system must address: \textit{context-awareness}, 
\textit{pro-activeness} of the system, and appropriate \textit{knowledge 
representation} are the main factors that impact on the quality of 
recommendations. Context-awareness becomes very relevant in software projects 
in which developers collaboratively work on shared resources. The 
pro-activeness of \rs  still demands further research due to the limited 
maturity of existing tools concerning such an aspect. A proactive approach 
should improve the accuracy of the recommendation by reducing the scope of the 
context. Finally, \rs should take into account more flexible representations of knowledge by considering new techniques, \eg Semantic Web analysis to 
make the system more transparent. 
%The authors also define the so-called \textit{recommendation landscape}, 
%consisting of two main dimensions: the \textit{knowledge sharing process} and 
%the \textit{recommended information}. The first dimension concerns manly the 
%awareness of the knowledge and the experience of capturing and sharing it. To 
%properly address these issues, shared knowledge must be supported using 
%interactive hints. Concerning recommended information, it characterizes items 
%referring to development activity \ie code example, artifacts or complete 
%tools 
%as well as the ones that describe the landscape of the project \ie accountable 
%of a certain feature, issues status, and the shared knowledge. 
After these considerations, the authors suggest that a combination of the 
context-awareness and the information provision can significantly improve the 
quality of the recommended items. In our work, we focus on \rs for mining OSS forges, \eg \GH, the Maven Central Repository, or Stack Overflow. Moreover, we tailor their design to satisfy different requirements of our industrial partners.

\smallskip
Gasparic and Janes \cite{gasparic_what_2016} provide a systematic literature 
review (SLR) on RSSE tools. The study aims to characterize RSSEs following the 
software engineers' perspective. In particular, the SLR focuses on the required 
inputs, on the benefits offered by the recommendation process, and on the 
required effort to provide the recommended items. As the topic is too broad, the 
authors put some constraints in building the research query. Recommendation 
systems that do not belong to the software engineering domain are excluded from 
the analysis. Among these, the study considers only stable tools, \ie not 
snapshot versions or reusable components. Additionally, the considered tools 
strictly support source code development and exclude other types of activities. 
Considering these boundaries, they conduct the review by analyzing every aspect 
of the process from the input extraction phase to the retrieved item. After an 
iterative process, 46 papers are selected for consideration. The major finding of 
the work is that most of the considered RSSEs use a reactive approach to 
provide recommendations. Reversely, proactive recommendations are not popular 
yet. The analyzed tools are focused 
on the development phase, but the testing phase is crucial in SE projects. 
Thus, future RSSEs should consider it as a possible application domain. Moreover, 
the examined context extraction techniques are not able to excerpt broader 
context, \ie they miss crucial elements useful to recommend valuable items. 
Concerning the presentation layer, tools should improve the explanation of the 
given recommendations and provide users with more information about their usage 
in concrete situations, \ie context-aware recommendations.  

\smallskip
%In \cite{ISINKAYE2015261}, the authors describe 
A taxonomy of recommendation systems as well as the possible phases involved in 
the recommendation process has been provided~\cite{ISINKAYE2015261}. 
The work describes recommendation process as an iterative cycle represented by 
three steps, \ie \textit{information gathering}, \textit{learning}, and 
\textit{recommendations delivery}. Information gathering involves the user's 
feedback collection at the beginning of the entire process after the system has 
issued the wanted recommendation. The user profile is considered to retrieve 
the proper items by looking at the user's needs in the presentation. Concerning 
information gathering, the authors classify possible feedback into 
\textit{explicit}, \textit{implicit} and \textit{hybrid}. Explicit feedback is given directly by the user, \ie through ratings which contain accurate 
information, but require effort to obtain. On the other hand, implicit feedback is inferred directly by the system without 
involving the user. However, they are less accurate than the explicit ones. The 
hybrid feedback is a combination of the previous two techniques. Typical 
implementations use the inferred data to check the feedback given by the user 
or allow the user to provide only a subset of information. Then, the learning 
phase employs them to excerpt the user's characteristics and to build a custom 
profile. 
Overall, this study~\cite{ISINKAYE2015261} highlights the 
contribution of different techniques as well as their strengths and weakness 
considering several factors, \ie availability of the meta-data, user ratings, 
and the learner model, to name a few. In the present paper, we 
conceptualized a novel taxonomy for the main design features for \rs in 
software engineering. Such a taxonomy is intended to provide system designers 
with the most pertinent technical details for their problems, \ie tailoring the 
design to satisfy the requirements imposed by use-case partners. 

\smallskip
%In \cite{laser_software_2015}, the authors show 
An extensive guideline that deals with challenges, issues, and basic blocks related to the development of RSSEs has been provided~\cite{laser_software_2015} to summarize well-established practices to show all the phases needed for the development of such systems. The first step involves \textit{problem framing}, \ie the identification of context, the tasks to be completed, and the target users of the recommended items. The context 
represents the development environment which brings plenty of information about 
the current task, which is the issue or functionality that the user is 
addressing.  Finally, the target user determines what kind of recommended items 
have to be provided and when: a novice developer's needs are profoundly 
different from those of expert ones. Thus, the final recommendation items might 
be completely different according to their experience level. 
Considering these aspects, the authors grouped RSSEs into four main 
categories:

\begin{itemize}
	
	\item \textit{Hotspot recommender}: it provides recommendations about 
	methods and classes that belong to the current context;
	\item \textit{Navigation recommender}: it suggests locations where the 
	developer can find hints related to the current task;
	\item \textit{Snippet recommender}: it produces snippets of code related to 
	the developer's context;
	\item \textit{Documentation recommender}: it aggregates posts coming from 
	websites and A\&Q forums to enhance the documentation of the APIs of 
	interest.
	
\end{itemize}

Moving to the \textit{recommendation process}, input sources must be handled to 
capture the developer's context. Due to their heterogeneity, RSSEs can employ 
different strategies and choose the most suitable for a certain context, such as static analysis, user feedback, structuring or destructuring techniques. 
This phase is usually followed by a preprocessing phase, in which data is rearranged for the following steps. Then, a recommendation algorithm is 
performed to obtain the recommendation item. According to the study~\cite{laser_software_2015}, recommendation algorithms fall in one of the 
following classes: 
\begin{itemize}
	\item \textit{Heuristic approaches} make effort on the implementation and usually derive from empirical evaluation;
	\item \textit{Data mining and machine learning techniques} are adopted when 
	a large amount of data are available, by exploiting different algorithm and 
	models;
	\item \textit{Collaborative filtering} typically employs user-item matrices to filter data and find similar items.
\end{itemize}

\noindent
The outcome of the recommendation algorithm is delivered to the target user in 
the \textit{presenting recommendation} phase. It is characterized by the level 
of interaction with the target user, which defines if the RSSE is proactive or 
reactive.

\smallskip
The work presented in this paper shares several similarities with the studies 
previously outlined. However, by leveraging the experiences we developed in the 
context of the \projectName project, the presented  challenges and lessons 
learned can complement the previous attempts of conceptualizing \rs with the 
aim of identifying their strengths and limitations when being applied in the 
context of software development.

%% file: src/Conclusions.tex
Developing complex software systems by reusing existing open source 
components is a challenging task.  In the EU \projectName project we worked on dealing with such a problem %in 
%the \projectName project 
by conceiving several \rs to meet the needs identified by six use-case 
partners working on different domains including IoT, multi-sector IT 
services, API co-evolution, software analytics, software quality assurance, 
and OSS forges.

In this paper, we presented an experience report on the various \rs that have 
been developed in \projectName. We attempt to share with the community the main challenges we had to 
overcome as well as the corresponding lessons during the three different phases to build 
and evaluate \rs, %concerning %we performed to conceive them 
	\ie requirement, development, and evaluation.

%a comprehensive review to 

%Comment 2.4: I would recommend to extend Section 6 in order to make it more 
%concrete, so that other researchers can better understand what are the 
%specific 
%actionable items coming from this paper. In other words, this must not only be 
%an experience report, but also something that the research community can build 
%upon.
%\revised{Conducting the project provides us with precious experience. In the 
%first place, we faced with challenges which emerge from different \rs.}

Being focused on heterogeneous \rs allowed us to garner many useful 
experiences and learn important lessons. In the first place, the process yields 
up a list of actionable items when designing and implementing \rs , namely: 
%related to: 
\textit{(i)} the skepticism that final users can have especially at the early 
stages of the development and usage of the proposed \rs; \textit{(ii)} 
difficulties in retrieving and creating datasets to be used both for training 
and evaluation purposes; \textit{(iii)} criticalities related to the selection 
of baselines for evaluation especially when the related tools are no longer 
available; \textit{(iv)} the variety of evaluation approaches and metrics that 
can be employed to assess the strengths and limitations of the conceived 
\rs.

%In this respect, Figure 10 (now Figure 3) is o

\emph{The first contribution} of our work is a taxonomy of the 
main design features for \rs in software engineering. We believe that such a 
taxonomy would come in handy for those who perform a fresh start on 
investigating which techniques are most suitable for their problems, \ie 
tailoring their design to meet the requirements imposed by industrial use-case 
partners. % To the best of our knowledge, no work has ever discussed such an 
%issue before.
%
%Moreover, %while developing the \projectName \rs, 
	\emph{The second contribution} of the paper is a benchmark consisting of the 
	challenges and lessons learned that might be useful for developers that 
	need to conceive new \rs and thus, have to encompass the three related 
	phases, \ie requirement elicitation, development, and evaluation. For instance, 
	through the evaluation of various systems, we realized that the selection 
	of suitable evaluation metrics helps shed light on the performance traits 
	that cannot be revealed by conventional indicators. %the overall 
	%performance. 
	In particular, while precision and recall represent the right choices for 
	assessing the quality of the proposed \rs, additional metrics typically 
	used in entirely different domains like sales diversity, and serendipity 
	can also be useful for %quality indicators when 
	studying a system's performance. Among others, the ability to recommend 
	items in the long tail is important: we can recommend items even when they 
	are extremely unpopular since each is used by a small number of projects. 
	However, they turn out to be useful as all of them match those stored as 
	ground-truth. This implies that the novelty of a ranked list is important: 
	a system should be able to recommend libraries that are \emph{novel} 
	\cite{Castells_noveltyand}, \ie those that have been rarely seen.  
	%interesting to apply

For future work, we plan to consolidate the lessons learned by applying in the 
Model Driven Engineering domain the techniques and tools we developed in 
\projectName. Moreover, we are also working on a low-code 
infrastructure to support the development of \rs. In particular, by relying on 
the presented taxonomy, we developed a metamodel to represent and manage the 
peculiar components of \rs \cite{10.1145/3417990.3420202}. Dedicated supporting 
tools are also under development to enable citizen developers to easily model 
and build their custom \rs. The results obtained so far are encouraging even 
though there is still significant work to be done to enable the development of 
\rs in a low-code manner.

%% file: main.bbl
\begin{thebibliography}{10}
\providecommand{\url}[1]{{#1}}
\providecommand{\urlprefix}{URL }
\expandafter\ifx\csname urlstyle\endcsname\relax
  \providecommand{\doi}[1]{DOI~\discretionary{}{}{}#1}\else
  \providecommand{\doi}{DOI~\discretionary{}{}{}\begingroup
  \urlstyle{rm}\Url}\fi

\bibitem{Alreshedy2018SCCAC}
Alreshedy, K., Dharmaretnam, D., German, D.M., Srinivasan, V., Gulliver, T.A.:
  {SCC}: {Automatic} classification of code snippets.
\newblock CoRR \textbf{abs/1809.07945} (2018)

\bibitem{7070485}
Basten, B., Hills, M., Klint, P., Landman, D., Shahi, A., Steindorfer, M.J.,
  Vinju, J.J.: M3: {A} general model for code analytics in rascal.
\newblock In: 2015 {IEEE} 1st international workshop on software analytics
  ({SWAN}), pp. 25--28 (2015)

\bibitem{BellogiN:2013:CSH:2397740.2398191}
Bellogín, A., Cantador, I., Castells, P.: A comparative study of heterogeneous
  item recommendations in social systems.
\newblock Inf. Sci. \textbf{221}, 142--169 (2013)

\bibitem{Blondel:2004:MSG:1035533.1035557}
Blondel, V.D., Gajardo, A., Heymans, M., Senellart, P., Dooren, P.V.: A measure
  of similarity between graph vertices: {Applications} to synonym extraction
  and web searching.
\newblock SIAM Review \textbf{46}(4), 647--666 (2004)

\bibitem{Bobadilla2013109}
Bobadilla, J., Ortega, F., Hernando, A., Gutiérrez, A.: Recommender systems
  survey.
\newblock Knowledge-Based Systems \textbf{46}, 109 -- 132 (2013)

\bibitem{borges_whats_2018}
Borges, H., Valente, M.T.: What's in a {GitHub} {Star}? {Understanding}
  {Repository} {Starring} {Practices} in a {Social} {Coding} {Platform}.
\newblock Journal of Systems and Software \textbf{146}, 112--129 (2018).
\newblock \doi{10.1016/j.jss.2018.09.016}.
\newblock \urlprefix\url{http://arxiv.org/abs/1811.07643}.
\newblock ArXiv: 1811.07643

\bibitem{Bruch:2008:ERS:1454247.1454254}
Bruch, M., Schäfer, T., Mezini, M.: On evaluating recommender systems for
  {API} usages.
\newblock In: Proceedings of the 2008 international workshop on recommendation
  systems for software engineering, {RSSE} '08, pp. 16--20. ACM, New York, NY,
  USA (2008)

\bibitem{Castells_noveltyand}
Castells, P., Vargas, S., Wang, J.: Novelty and diversity metrics for
  recommender systems: Choice, discovery and relevance.
\newblock In: International Workshop on Diversity in Document Retrieval (DDR
  2011) at the 33rd European Conference on Information Retrieval (ECIR 2011).
  Dublin, Ireland (2011).
\newblock \urlprefix\url{http://ir.ii.uam.es/rim3/publications/ddr11.pdf}

\bibitem{7887704}
{Cosentino}, V., {Cánovas Izquierdo}, J.L., {Cabot}, J.: A systematic mapping
  study of software development with github.
\newblock IEEE Access \textbf{5}, 7173--7192 (2017).
\newblock \doi{10.1109/ACCESS.2017.2682323}

\bibitem{Czarnecki02}
Czarnecki, K.: Domain Engineering, pp. 433--444.
\newblock American Cancer Society (2002).
\newblock \doi{10.1002/0471028959.sof095}.
\newblock
  \urlprefix\url{https://onlinelibrary.wiley.com/doi/abs/10.1002/0471028959.sof095}

\bibitem{Dagenais:2010:MNS:1806799.1806842}
Dagenais, B., Ossher, H., Bellamy, R.K.E., Robillard, M.P., de~Vries, J.P.:
  Moving into a new software project landscape.
\newblock In: Proceedings of the {32Nd} {ACM}/{IEEE} international conference
  on software engineering - volume 1, {ICSE} '10, pp. 275--284. ACM, New York,
  NY, USA (2010)

\bibitem{Davis:2006:RPR:1143844.1143874}
Davis, J., Goadrich, M.: The relationship between precision-recall and {ROC}
  curves.
\newblock In: Proceedings of the 23rd international conference on machine
  learning, {ICML} '06, pp. 233--240. ACM, New York, NY, USA (2006)

\bibitem{SH}
Di~Cosmo, R., Zacchiroli, S.: {Software Heritage: Why and How to Preserve
  Software Source Code}.
\newblock In: 14th International Conference on Digital Preservation, pp. 1--10.
  Kyoto (2017)

\bibitem{DiNoia:2012:LOD:2362499.2362501}
Di~Noia, T., Mirizzi, R., Ostuni, V.C., Romito, D., Zanker, M.: Linked open
  data to support content-based recommender systems.
\newblock In: Proceedings of the 8th international conference on semantic
  systems, I-semantics '12, pp. 1--8. ACM, New York, NY, USA (2012)

\bibitem{10.1145/3382494.3410690}
Di~Rocco, J., Di~Ruscio, D., Di~Sipio, C., Nguyen, P., Rubei, R.: Topfilter: An
  approach to recommend relevant github topics.
\newblock In: Proceedings of the 14th ACM / IEEE International Symposium on
  Empirical Software Engineering and Measurement (ESEM), ESEM '20. Association
  for Computing Machinery, New York, NY, USA (2020).
\newblock \doi{10.1145/3382494.3410690}

\bibitem{10.1145/3417990.3420202}
Di~Sipio, C., Di~Ruscio, D., Nguyen, P.T.: Democratizing the development of
  recommender systems by means of low-code platforms.
\newblock In: Proceedings of the 23rd ACM/IEEE International Conference on
  Model Driven Engineering Languages and Systems: Companion Proceedings, MODELS
  '20. Association for Computing Machinery, New York, NY, USA (2020).
\newblock \doi{10.1145/3417990.3420202}

\bibitem{10.1145/3383219.3383227}
Di~Sipio, C., Rubei, R., {Di Ruscio}, D., Nguyen, P.T.: {Using a Multinomial
  Na\"ive Bayesian (MNB) Network to Automatically Recommend Topics for GitHub
  Repositories}.
\newblock In: Proceedings of the 24th International Conference on Evaluation
  and Assessment in Software Engineering, EASE2020, Trondheim, Norway, April
  15-17, 2020, EASE’20, pp. 24--34. {ACM} (2020).
\newblock \doi{10.1145/3383219.3383227}

\bibitem{Fowkes:2016:PPA:2950290.2950319}
Fowkes, J., Sutton, C.: Parameter-free probabilistic {API} mining across
  {GitHub}.
\newblock In: Proceedings of the 2016 24th {ACM} {SIGSOFT} international
  symposium on foundations of software engineering, {FSE} 2016, pp. 254--265.
  ACM, New York, NY, USA (2016)

\bibitem{ganesan_topic_2017}
Ganesan, K.: Topic {Suggestions} for {Millions} of {Repositories} - {The}
  {GitHub} {Blog} (2017).
\newblock \urlprefix\url{https://github.blog/2017-07-31-topics/}

\bibitem{10.1109/APSEC.2004.69}
Garg, P.K., Kawaguchi, S., Matsushita, M., Inoue, K.: {MUDABlue}: {An}
  automatic categorization system for open source repositories.
\newblock 2013 20th Asia-Pacific Software Engineering Conference (APSEC) pp.
  184--193 (2004)

\bibitem{gasparic_what_2016}
Gasparic, M., Janes, A.: What recommendation systems for software engineering
  recommend: {A} systematic literature review.
\newblock Journal of Systems and Software \textbf{113}, 101--113 (2016).
\newblock \doi{10.1016/j.jss.2015.11.036}.
\newblock
  \urlprefix\url{https://linkinghub.elsevier.com/retrieve/pii/S0164121215002605}

\bibitem{10.1145/1864708.1864761}
Ge, M., Delgado-Battenfeld, C., Jannach, D.: Beyond accuracy: Evaluating
  recommender systems by coverage and serendipity.
\newblock In: Proceedings of the Fourth ACM Conference on Recommender Systems,
  RecSys ’10, p. 257–260. Association for Computing Machinery, New York,
  NY, USA (2010).
\newblock \doi{10.1145/1864708.1864761}.
\newblock \urlprefix\url{https://doi.org/10.1145/1864708.1864761}

\bibitem{Ghose2001}
Ghose, S., Lowengart, O.: Taste tests: {Impacts} of consumer perceptions and
  preferences on brand positioning strategies.
\newblock Journal of Targeting, Measurement and Analysis for Marketing
  \textbf{10}(1), 26--41 (2001)

\bibitem{Gomez-Uribe:2015:NRS:2869770.2843948}
Gomez-Uribe, C.A., Hunt, N.: The netflix recommender system: {Algorithms},
  business value, and innovation.
\newblock ACM Transactions on Management Information Systems \textbf{6}(4),
  13:1--13:19 (2015)

\bibitem{Gousi13}
Gousios, G.: The ghtorrent dataset and tool suite.
\newblock In: Proceedings of the 10th Working Conference on Mining Software
  Repositories, MSR '13, pp. 233--236. IEEE Press, Piscataway, NJ, USA (2013).
\newblock \urlprefix\url{http://dl.acm.org/citation.cfm?id=2487085.2487132}

\bibitem{happel_potentials_2008}
Happel, H.J., Maalej, W.: Potentials and challenges of recommendation systems
  for software development.
\newblock In: Proceedings of the 2008 international workshop on
  {Recommendation} systems for software engineering - {RSSE} '08, p.~11. ACM
  Press, Atlanta, Georgia (2008)

\bibitem{holmes_strathcona_nodate}
Holmes, R., Walker, R.J., Murphy, G.C.: Strathcona {Example} {Recommendation}
  {Tool} p.~4 (2015)

\bibitem{ISINKAYE2015261}
Isinkaye, F., Folajimi, Y., Ojokoh, B.: Recommendation systems: {Principles},
  methods and evaluation.
\newblock Egyptian Informatics Journal \textbf{16}(3), 261 -- 273 (2015)

\bibitem{Jiang:2017:WDF:3042021.3042043}
Jiang, J., Lo, D., He, J., Xia, X., Kochhar, P.S., Zhang, L.: {Why and How
  Developers Fork What from Whom in GitHub}.
\newblock Empirical Software Engineering \textbf{22}(1), 547--578 (2017).
\newblock \doi{10.1007/s10664-016-9436-6}.
\newblock \urlprefix\url{https://doi.org/10.1007/s10664-016-9436-6}

\bibitem{Karlsson1995}
Karlsson, E.A. (ed.): Software Reuse: A Holistic Approach.
\newblock John Wiley \& Sons, Inc., New York, NY, USA (1995)

\bibitem{kendall1948rank}
Kendall, M.G.: {A New Measure of Rank Correlation}.
\newblock Biometrika \textbf{30}(1/2), 81--93 (1938).
\newblock \urlprefix\url{http://www.jstor.org/stable/2332226}

\bibitem{hutchison_multinomial_2004}
Kibriya, A.M., Frank, E., Pfahringer, B., Holmes, G.: Multinomial naive bayes
  for text categorization revisited.
\newblock In: G.I. Webb, X.~Yu (eds.) AI 2004: Advances in Artificial
  Intelligence, pp. 488--499. Springer Berlin Heidelberg, Berlin, Heidelberg
  (2005)

\bibitem{laser_software_2015}
{LASER}, {LASER}: Software engineering: international summer schools, {LASER}
  2013-2014, {Elba}, {Italy}: revised tutorial lectures.
\newblock No. 8987 in Lecture notes in computer science {Programming} and
  software engineering. Springer, [Cham] Heidelberg (2015)

\bibitem{Linden:2003:ARI:642462.642471}
Linden, G., Smith, B., York, J.: Amazon.{Com} recommendations: {Item}-to-item
  collaborative filtering.
\newblock IEEE Internet Computing \textbf{7}(1), 76--80 (2003)

\bibitem{DBLP:conf/kbse/LvZLWZZ15}
Lv, F., Zhang, H., Lou, J.G., Wang, S., Zhang, D., Zhao, J.: {CodeHow}:
  {Effective} code search based on {API} understanding and extended boolean
  model ({E}).
\newblock In: 30th {IEEE}/{ACM} international conference on automated software
  engineering, {ASE} 2015, lincoln, {NE}, {USA}, november 9-13, 2015, pp.
  260--270 (2015)

\bibitem{maki_context}
Maki, S., Kpodjedo, S., Boussaidi, G.E.: Context {Extraction} in
  {Recommendation} {Systems} in {Software} {Engineering}: {A} {Preliminary}
  {Survey} p.~10 (2015)

\bibitem{McMillan:2012:DSS:2337223.2337267}
McMillan, C., Grechanik, M., Poshyvanyk, D.: Detecting similar software
  applications.
\newblock In: Proceedings of the 34th international conference on software
  engineering, {ICSE} '12, pp. 364--374. IEEE Press, Piscataway, NJ, USA (2012)

\bibitem{McMillan:2010:RSC:1808920.1808925}
McMillan, C., Poshyvanyk, D., Grechanik, M.: Recommending source code examples
  via {API} call usages and documentation.
\newblock In: Proceedings of the {2Nd} international workshop on recommendation
  systems for software engineering, {RSSE} '10, pp. 21--25. ACM, New York, NY,
  USA (2010)

\bibitem{Moreno:2015:IUT:2818754.2818860}
Moreno, L., Bavota, G., Di~Penta, M., Oliveto, R., Marcus, A.: How can {I} use
  this method?
\newblock In: 37th international conference on software engineering, pp.
  880--890. IEEE, Piscataway (2015)

\bibitem{10.1145/2740908.2742141}
Nguyen, P., Tomeo, P., Di~Noia, T., Di~Sciascio, E.: {An Evaluation of SimRank
  and Personalized PageRank to Build a Recommender System for the Web of Data}.
\newblock In: Proceedings of the 24th International Conference on World Wide
  Web, WWW '15 Companion, p. 1477–1482. Association for Computing Machinery,
  New York, NY, USA (2015).
\newblock \doi{10.1145/2740908.2742141}.
\newblock
  \urlprefix\url{https://doi-org.univaq.clas.cineca.it/10.1145/2740908.2742141}

\bibitem{DBLP:conf/iir/NguyenRR18}
Nguyen, P.T., Di~Rocco, J., Di~Ruscio, D.: Mining software repositories to
  support {OSS} developers: {A} recommender systems approach.
\newblock In: Proceedings of the 9th italian information retrieval workshop,
  rome, italy, may, 28-30, 2018. (2018)

\bibitem{Nguyen:2019:JSS:CrossRec}
Nguyen, P.T., {Di Rocco}, J., {Di Ruscio}, D., {Di Penta}, M.: {CrossRec:
  Supporting Software Developers by Recommending Third-party Libraries}.
\newblock Journal of Systems and Software p. 110460 (2019).
\newblock \doi{https://doi.org/10.1016/j.jss.2019.110460}.
\newblock
  \urlprefix\url{http://www.sciencedirect.com/science/article/pii/S0164121219302341}

\bibitem{Nguyen:2019:FRS:3339505.3339636}
Nguyen, P.T., Di~Rocco, J., Di~Ruscio, D., Ochoa, L., Degueule, T., Di~Penta,
  M.: {FOCUS}: {A} recommender system for mining {API} function calls and usage
  patterns.
\newblock In: Proceedings of the 41st international conference on software
  engineering, {ICSE} '19, pp. 1050--1060. IEEE Press, Piscataway, NJ, USA
  (2019)

\bibitem{9359479}
{Nguyen}, P.T., {Di Rocco}, J., {Di Sipio}, C., {Di Ruscio}, D., {Di Penta},
  M.: Recommending api function calls and code snippets to support software
  development.
\newblock IEEE Transactions on Software Engineering pp. 1--1 (2021).
\newblock \doi{10.1109/TSE.2021.3059907}

\bibitem{8498236}
Nguyen, P.T., Di~Rocco, J., Rubei, R., Di~Ruscio, D.: {CrossSim}: {Exploiting}
  mutual relationships to detect similar {OSS} projects.
\newblock In: 2018 44th euromicro conference on software engineering and
  advanced applications ({SEAA}), pp. 388--395 (2018)

\bibitem{Nguyen:2019:JSS:CrossSim}
Nguyen, P.T., {Di Rocco}, J., Rubei, R., {Di Ruscio}, D.: An automated approach
  to assess the similarity of {GitHub} repositories.
\newblock Software Quality Journal  (2020).
\newblock \doi{10.1007/s11219-019-09483-0}.
\newblock \urlprefix\url{https://doi.org/10.1007%2Fs11219-019-09483-0}

\bibitem{Nguyen:2015:CRV:2942298.2942305}
Nguyen, P.T., Tomeo, P., Di~Noia, T., Di~Sciascio, E.: Content-based
  recommendations via {DBpedia} and freebase: {A} case study in the music
  domain.
\newblock In: Proceedings of the 14th international conference on the semantic
  web - {ISWC} 2015 - volume 9366, pp. 605--621. Springer-Verlag New York,
  Inc., New York, NY, USA (2015)

\bibitem{niuAPIUsagePattern2017}
Niu, H., Keivanloo, I., Zou, Y.: {API} usage pattern recommendation for
  software development.
\newblock Journal of Systems and Software \textbf{129}, 127--139 (2017)

\bibitem{DBLP:conf/rweb/NoiaO15}
Noia, T.D., Ostuni, V.C.: Recommender systems and linked open data.
\newblock In: W.~Faber, A.~Paschke (eds.) Reasoning Web. Web Logic Rules - 11th
  International Summer School 2015, Berlin, Germany, July 31 - August 4, 2015,
  Tutorial Lectures, \emph{Lecture Notes in Computer Science}, vol. 9203, pp.
  88--113. Springer (2015).
\newblock \doi{10.1007/978-3-319-21768-0\_4}.
\newblock \urlprefix\url{https://doi.org/10.1007/978-3-319-21768-0\_4}

\bibitem{Ouni:2017:SSL:3032135.3032325}
Ouni, A., Kula, R.G., Kessentini, M., Ishio, T., German, D.M., Inoue, K.:
  Search-based software library recommendation using multi-objective
  optimization.
\newblock Inf. Softw. Technol. \textbf{83}(C), 55--75 (2017)

\bibitem{pakdeetrakulwong_recommendation_2014}
Pakdeetrakulwong, U., Wongthongtham, P., Siricharoen, W.V.: Recommendation
  systems for software engineering: {A} survey from software development life
  cycle phase perspective.
\newblock pp. 137--142. IEEE (2014)

\bibitem{doi:10.1108/13522750810879048}
Pettigrew, S., Charters, S.: Tasting as a projective technique.
\newblock Qualitative Market Research: An International Journal \textbf{11}(3),
  331--343 (2008)

\bibitem{ponzanelli_prompter:_2016}
Ponzanelli, L., Bavota, G., Di~Penta, M., Oliveto, R., Lanza, M.: Prompter:
  {Turning} the {IDE} into a self-confident programming assistant.
\newblock Empirical Software Engineering \textbf{21}(5), 2190--2231 (2016).
\newblock \doi{10.1007/s10664-015-9397-1}.
\newblock \urlprefix\url{http://link.springer.com/10.1007/s10664-015-9397-1}

\bibitem{DBLP:journals/ese/PonzanelliBPOL16}
Ponzanelli, L., Bavota, G., Penta, M.D., Oliveto, R., Lanza, M.: Prompter -
  {Turning} the {IDE} into a self-confident programming assistant.
\newblock Empirical Software Engineering \textbf{21}(5), 2190--2231 (2016)

\bibitem{DBLP:conf/laser/ProkschBM14}
Proksch, S., Bauer, V., Murphy, G.C.: How to build a recommendation system for
  software engineering.
\newblock In: B.~Meyer, M.~Nordio (eds.) Software Engineering - International
  Summer Schools, {LASER} 2013-2014, Elba, Italy, Revised Tutorial Lectures,
  \emph{Lecture Notes in Computer Science}, vol. 8987, pp. 1--42. Springer
  (2014).
\newblock \doi{10.1007/978-3-319-28406-4\_1}.
\newblock \urlprefix\url{https://doi.org/10.1007/978-3-319-28406-4\_1}

\bibitem{tubiblio77729}
Proksch, S., Bauer, V., Murphy, G.C.: How to build a recommendation system for
  software engineering.
\newblock In: B.~Meyer, M.~Nordio (eds.) Advances in the theory and practice of
  software engineering - LASER 2013-2014, \emph{LNCS}, vol. 8987, pp. 1--42.
  Springer (2015).
\newblock \urlprefix\url{http://tubiblio.ulb.tu-darmstadt.de/77729/}

\bibitem{Robillard:2013:AAP:2498733.2498776}
Robillard, M.P., Bodden, E., Kawrykow, D., Mezini, M., Ratchford, T.: Automated
  {API} property inference techniques.
\newblock IEEE Transactions on Software Engineering \textbf{39}(5), 613--637
  (2013)

\bibitem{robillard_recommendation_2014}
Robillard, M.P., Maalej, W., Walker, R.J., Zimmermann, T. (eds.):
  Recommendation {Systems} in {Software} {Engineering}.
\newblock Springer Berlin Heidelberg, Berlin, Heidelberg (2014)

\bibitem{RUBEI2020106367}
Rubei, R., {Di Sipio}, C., Nguyen, P.T., {Di Rocco}, J., {Di Ruscio}, D.:
  {PostFinder: Mining Stack Overflow posts to support software developers}.
\newblock Information and Software Technology \textbf{127}, 106367 (2020).
\newblock \doi{https://doi.org/10.1016/j.infsof.2020.106367}.
\newblock
  \urlprefix\url{http://www.sciencedirect.com/science/article/pii/S0950584920301361}

\bibitem{SAIED2018164}
Saied, M.A., Ouni, A., Sahraoui, H., Kula, R.G., Inoue, K., Lo, D.: Improving
  reusability of software libraries through usage pattern mining.
\newblock Journal of Systems and Software \textbf{145}, 164 -- 179 (2018)

\bibitem{DBLP:journals/ijmir/SchedlZCDE18}
Schedl, M., Zamani, H., Chen, C., Deldjoo, Y., Elahi, M.: Current challenges
  and visions in music recommender systems research.
\newblock Int. J. Multim. Inf. Retr. \textbf{7}(2), 95--116 (2018).
\newblock \doi{10.1007/s13735-018-0154-2}.
\newblock \urlprefix\url{https://doi.org/10.1007/s13735-018-0154-2}

\bibitem{spearman1904proof}
Spearman, C.: The proof and measurement of association between two things.
\newblock The American journal of psychology \textbf{15}(1), 72--101 (1904)

\bibitem{SS04}
Spinellis, D., Szyperski, C.: How is open source affecting software
  development?
\newblock IEEE Software \textbf{21}(1), 28--33 (2004)

\bibitem{6671293}
Thung, F., Lo, D., Lawall, J.: Automated library recommendation.
\newblock In: 2013 20th working conf. on reverse engineering ({WCRE}), pp.
  182--191 (2013)

\bibitem{Vargas:2011:RRN:2043932.2043955}
Vargas, S., Castells, P.: Rank and relevance in novelty and diversity metrics
  for recommender systems.
\newblock In: Proceedings of the fifth {ACM} conference on recommender systems,
  {RecSys} '11, pp. 109--116. ACM, New York, NY, USA (2011)

\bibitem{Vargas_sales_diversity_14}
Vargas, S., Castells, P.: Improving sales diversity by recommending users to
  items.
\newblock In: Eighth {ACM} conference on recommender systems, {RecSys} '14,
  foster city, silicon valley, {CA}, {USA} - october 06 - 10, 2014, pp.
  145--152 (2014)

\bibitem{Wang2013Mining}
Wang, J., Dang, Y., Zhang, H., Chen, K., Xie, T., Zhang, D.: {Mining Succinct
  and High-coverage API Usage Patterns from Source Code}.
\newblock In: 10th Working Conference on Mining Software Repositories, pp.
  319--328. IEEE, Piscataway (2013).
\newblock \doi{10.1109/MSR.2013.6624045}

\bibitem{wolpert_no_1997}
Wolpert, D., Macready, W.: No free lunch theorems for optimization.
\newblock IEEE Transactions on Evolutionary Computation \textbf{1}(1), 67--82
  (1997).
\newblock \doi{10.1109/4235.585893}.
\newblock \urlprefix\url{http://ieeexplore.ieee.org/document/585893/}

\bibitem{WONG20152839}
Wong, T.T.: {Performance Evaluation of Classification Algorithms by K-fold and
  Leave-one-out Cross Validation}.
\newblock Pattern Recognition \textbf{48}(9), 2839--2846 (2015).
\newblock \doi{10.1016/j.patcog.2015.03.009}

\bibitem{DBLP:conf/recsys/WuSCTP14}
Wu, L., Shah, S., Choi, S., Tiwari, M., Posse, C.: The browsemaps:
  {Collaborative} filtering at {LinkedIn}.
\newblock In: {RSWeb}@{RecSys}, \emph{{CEUR} workshop proceedings}, vol. 1271.
  CEUR-WS.org (2014)

\bibitem{10.1109/SANER.2017.7884605}
Zhang, Y., Lo, D., Kochhar, P.S., Xia, X., Li, Q., Sun, J.: Detecting similar
  repositories on {GitHub}.
\newblock 2017 IEEE 24th International Conference on Software Analysis,
  Evolution and Reengineering (SANER) \textbf{00}, 13--23 (2017)

\bibitem{DBLP:conf/aics/ZhengPL18}
Zheng, M., Pan, X., Lillis, D.: {CodEX}: {Source} code plagiarism detection
  based on abstract syntax tree.
\newblock In: Proceedings for the 26th {AIAI} irish conference on artificial
  intelligence and cognitive science trinity college dublin, dublin, ireland,
  december 6-7th, 2018., pp. 362--373 (2018)

\bibitem{10.1007/978-3-642-03013-0_15}
Zhong, H., Xie, T., Zhang, L., Pei, J., Mei, H.: {MAPO}: {Mining} and
  recommending {API} usage patterns.
\newblock In: 23rd european conference on object-oriented programming, pp.
  318--343. Springer, Berlin, Heidelberg (2009)

\end{thebibliography}
